\documentclass[lettersize,journal]{IEEEtran}
\usepackage{amsmath,amsfonts}
\usepackage{algorithmic}
\usepackage{array}
\usepackage[caption=false,font=normalsize,labelfont=sf,textfont=sf]{subfig}
\usepackage{textcomp}
\usepackage{stfloats}
\usepackage{url}
\usepackage{verbatim}
\usepackage{graphicx}
\def\BibTeX{{\rm B\kern-.05em{\sc i\kern-.025em b}\kern-.08em
    T\kern-.1667em\lower.7ex\hbox{E}\kern-.125emX}}
\usepackage{balance}

\usepackage{amsmath}
\usepackage{amssymb}
\usepackage{mathtools}
\usepackage{bm}
\usepackage[hidelinks]{hyperref}
\usepackage{booktabs}
\usepackage{lipsum}
\usepackage{xcolor}
\usepackage{doi}
\usepackage[numbers,compress]{natbib}

\definecolor{ML_c1}{rgb}{0, 0.447, 0.741}
\definecolor{ML_c2}{rgb}{0.85, 0.325, 0.098}
\definecolor{ML_c3}{rgb}{0.929, 0.694, 0.125}
\definecolor{ML_c4}{rgb}{0.494, 0.184, 0.556}

\usepackage{amsthm}
\usepackage{aliascnt}                       % To allow proper referencing of theorems using \autoref
\theoremstyle{remark}% Non-bold, italics
\newtheorem{thm}{Theorem}
\newtheorem{lem}{Lemma}
\newtheorem{cor}{Corollary}

\newtheorem{defn}{Definition}
\newtheorem{rem}{Remark} 
\newtheorem{ass}{Assumption}

\newtheorem{alg}{Algorithm}
\long\def\@makealgocaption#1#2{\vskip 2ex \small
  \hbox to \hsize{\parbox[t]{\hsize}{{\bfseries #1.} #2}}}
\newcounter{algorithm}
\def\thealgorithm{\@arabic\c@algorithm}
\def\fps@algorithm{tbp}
\def\ftype@algorithm{4}
\def\ext@algorithm{lof}
\def\fnum@algorithm{Algorithm \thealgorithm}
\def\equationautorefname~#1\null{%
  (#1)\null
}

\usepackage[outline]{contour}
\newcommand*\groundconn{\includegraphics{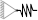}}
\newcommand*\IOsymb{\includegraphics{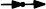}}
\newcommand*\errboundbox{\raisebox{-.2mm}{\includegraphics[scale =0.28]{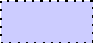}}}
\newcommand*\bbDelta{\includegraphics[scale =0.7]{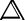}}
\newcommand*\bbOmega{\includegraphics{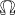}}
\newcommand*\bbLambda{\includegraphics[scale =0.7]{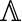}}

\newcommand\shdots{\makebox[1em][c]{.\hfil.\hfil.}}

\usepackage{tikz}

\newcommand\FOMline{\hspace{-.7mm}
\begin{tikzpicture}
    \draw[line width=0.5mm, black ] (0,0) -- (0.3,0) node {} ;
\end{tikzpicture}\hspace{-.7mm}}
\newcommand\ssBRline{\hspace{-.7mm}
\begin{tikzpicture}
    \draw[line width=0.5mm, ML_c3 ] (0,0) -- (0.05,0) node{} ;
    \draw[line width=0.5mm, ML_c3] (0.1,0) -- (0.15,0) node {} ;
    \draw[line width=0.5mm, ML_c3 ] (0.2,0) -- (0.25,0)node {}  ;
    \draw[line width=0.5mm, ML_c3 ] (0.3,0) -- (0.35,0) node {} ;
\end{tikzpicture}\hspace{-.7mm}}
\newcommand\ISBRline{\hspace{-.7mm}
\begin{tikzpicture}
    \draw[line width=0.5mm, ML_c1 ] (0,0) -- (0.1,0) node {} ;
    \draw[line width=0.5mm, ML_c1] (0.15,0) -- (0.2,0) node {} ;
    \draw[line width=0.5mm, ML_c1 ] (0.25,0) -- (0.35,0)node {}  ;
\end{tikzpicture}\hspace{-.7mm}}
\newcommand\aCLBRline{\hspace{-.7mm}
\begin{tikzpicture}
    \draw[line width=0.5mm, ML_c2 ] (0,0) -- (0.12,0) node {} ;
    \draw[line width=0.5mm, ML_c2] (0.24,0) -- (0.36,0) node {} ;
\end{tikzpicture}\hspace{-.7mm}}

\newcommand{\red}{}

\DeclareMathOperator\trace{tr}
\DeclareMathOperator\diag{diag}

\newcommand{\R}{\mathbb{R}}
\newcommand{\C}{\mathbb{C}}

\newcommand{\Lnrm}[1]{{\ensuremath{\mathcal{L}_{#1}}}}
\newcommand{\Hnrm}[1]{{\ensuremath{\mathcal{H}_{#1}}}}
\newcommand{\RH}{\mathcal{R}\Hnrm{\infty}}

\newcommand{\llft}{\mathcal{F}_l}       % Lower LFT
\newcommand{\ulft}{\mathcal{F}_u}       % Upper LFT

\newcommand{\sys}{\Sigma}          % System model
\newcommand{\sysr}{\hat{\sys}}          % Reduced system

\newcommand{\env}{E}               % Environment model
\newcommand{\envr}{\hat{\env}}        % Reduced environment

\newcommand{\fnv}{F}               % Extended environment
\newcommand{\fnvr}{\hat{\fnv}}          % Extended reduced environment
\newcommand{\ee}{\Lambda_E}  
\newcommand{\ees}[1]{\Lambda_{E,#1}}             
\newcommand{\ef}{\Lambda_F}        
\newcommand{\efs}[1]{\Lambda_{F,#1}}         
\newcommand{\eft}{\tilde{\Lambda}_F}             
\newcommand{\ec}{\Lambda_C}

\newcommand{\des}[1]{\Delta_{E,#1}}

\newcommand{\dft}{\tilde{\Delta}_F}             
\newcommand{\dc}{\Delta_C}       

\newcommand{\ue}{\bbLambda} 
    
\newcommand{\uees}[1]{\ue_{E,#1}}            
                
\newcommand{\ueft}{\tilde{\ue}_F}             
             
\newcommand{\uec}{\ue_C}    

\newcommand{\ud}{\bbDelta} 
\newcommand{\sud}{\scalebox{0.7}{\ud}} % Small \ue for subscript

\newcommand{\eb}{\bar{\varepsilon}} 
\newcommand{\ebe}{\eb_E}    
            
\newcommand{\ebf}{\eb_F}

\newcommand{\cV}{\mathcal{V}}
\newcommand{\cW}{\mathcal{W}}
\newcommand{\bV}{\bar{V}}

\usepackage[export]{adjustbox}

\author{Luuk Poort, Lars A.L. Janssen, Bart Besselink, Rob H.B. Fey, Nathan van de Wouw
\thanks{Manuscript created November, 2024; This work is funded by Holland High Tech \textbar \ TKI HSTM via the PPS allowance scheme for public-private partnerships and ASML.\\
The authors Luuk Poort, Lars Janssen, Rob Fey and Nathan van de Wouw are with the Department of Mechanical Engineering, in the Dynamics \& Control group, Eindhoven University of Technology, 5600 MB Eindhoven, the Netherlands (e-mail: l.poort@tue.nl; l.a.l.janssen@tue.nl; r.h.b.fey@tue.nl; n.v.d.wouw@tue.nl).\\
Bart Besselink is with the Bernoulli Institute for Mathematics, Computer Science and Artificial Intelligence, University of Groningen, 9700 AK, Groningen, the Netherlands (e-mail: b.besselink@rug.nl).
}}
\title{Abstracted Model Reduction:\\ \LARGE{A General Framework for Efficient Interconnected System Reduction}}

\begin{document}
\setcitestyle{square}
\maketitle

%
%%%%%%%%%%%%%%%%%%%%%%%%%%%%%%%%%%%%%%%%%
%
\begin{abstract}
    This paper introduces the concept of abstracted model reduction: a framework to improve the tractability of structure-preserving methods for the complexity reduction of interconnected system models. To effectively reduce high-order, interconnected models, it is usually not sufficient to consider the subsystems separately. Instead, structure-preserving reduction methods should be employed, which consider the interconnected dynamics to select which subsystem dynamics to retain in reduction. However, structure-preserving methods are often not computationally tractable. To overcome this issue, we propose to connect each subsystem model to a low-order abstraction of its environment to reduce it both effectively and efficiently. By means of a high-fidelity structural-dynamics model from the lithography industry, we show, on the one hand, significantly increased accuracy with respect to standard subsystem reduction and, on the other hand, similar accuracy to direct application of expensive structure-preserving methods, while significantly reducing computational cost. Furthermore, we formulate a systematic approach to automatically determine sufficient abstraction and reduction orders to preserve stability and guarantee a given frequency-dependent error specification. We apply this approach to the lithography equipment use case and show that the environment model can indeed be reduced by over 80\% without significant loss in the accuracy of the reduced interconnected model.
\end{abstract}

\begin{IEEEkeywords}
Model Reduction, Interconnected Systems, Stability Preservation,  Accuracy Guarantee, Structural Dynamics
\end{IEEEkeywords}

%
%%%%%%%%%%%%%%%%%%%%%%%%%%%%%%%%%%%%%%%%%
%

\section{Introduction} \label{sec:intro}
\begin{figure}\hspace{15mm}
    \subfloat[\small\label{fig:SS_RGB}]{%
      \includegraphics[scale = 1]{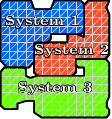}
    } \hfill
    \subfloat[\small\label{fig:SE_RGB}]{%
      \includegraphics[scale = 1]{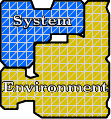}
    }\hspace{15mm}
    \caption{The interconnected system as a) an interconnection of three subsystems, and b) an interconnection of a single system and its environment.}
    \label{fig:SS-SE_RGB}
\end{figure}

Complex dynamical systems are often composed of interconnected subsystems, such as plant-controller feedback loops, multi-physical systems, networked cyber-physical systems, or assemblies of components in high-tech equipment. The collective of all subsystem models constitutes the model of the interconnected system, as schematically visualized in \autoref{fig:SS_RGB}. Often, this interconnected system model is of such high order that dynamical analysis becomes computationally infeasible and model order reduction methods are required to approximate the high-order model by a low-order surrogate model. An illustrative scenario where this problem arises can be found in structural dynamics, such as in the design and analysis of lithography machines \cite{Dorosti2014FiniteControl}, which serves as the primary motivating case study for this research.

When reducing a model of interconnected subsystems, it is preferable to retain the interconnection structure to keep the modelling approach modular. In other words, the reduced interconnected system model should be constructed from reduced subsystem models. This modular reduction approach aligns with the typical design process in industry, where each subsystem is largely developed separately \cite{Baldwin2006ModularitySystems}. The combination of such modular design and model reduction allows for greater flexibility of individual design teams, preservation of the essential structure of the interconnected system and increased interpretability of the reduced system model. 

The most direct and resource-efficient approach to modular reduction involves individually reducing each subsystem model \cite{Reis2008ASystems}, i.e., in an ``open-loop'' sense, for which many standard reduction methods exist \cite{Besselink2013AControl,DeKlerk2008GeneralTechniques,Antoulas2005ApproximationSystems}. This approach is particularly practical because the computational cost of model reduction scales with the model order and the individual subsystem models are clearly of lower order than the interconnected model as a whole. Even though one has to reduce multiple lower-order subsystem models, this is usually still much cheaper than the reduction of one high-order interconnected system model. However, individual reduction of the separate subsystem models might not retain the dynamics required for the reduced, interconnected system model to accurately approximate its full-order model counterpart \cite{Reis2008ASystems,Sandberg2009}. 

To improve the accuracy of the reduced interconnected system model, a variety of methods \cite{Wortelboer1994FrequencyConfiguration,Vandendorpe2008ModelSystems,Poort2024Balancing-BasedSystems,Kessels2022Sensitivity-BasedReduction,Kim2018ADisplacement,Li2005Structure-PreservingFormulation,Li2005StructuredLMIs} have been developed to reduce the interconnected system model while preserving its interconnection structure. These methods effectively reduce the individual subsystem models in a ``closed-loop'' sense by considering the interconnected dynamics. Consequently, such \emph{structure-preserving} reduction methods successfully reduce the subsystem models to lower-order models that collectively constitute a low-order, interconnected model that accurately approximates the original, high-order model.

Unfortunately, for many complex engineering systems the interconnected system model is of such a high order, that many structure-preserving reduction methods become computationally infeasible. In such cases, only computationally highly efficient reduction methods remain viable.

One approach to address these computational constraints is to use approximate, iterative methods. For instance, in \cite{Villena2009BlockSystems}, the Gramians of the interconnected system model are approximated to reduce each subsystem model. However, as available structure-preserving reduction methods evaluate the interconnected model differently, i.e., not all use Gramians \cite{Reis2008ASystems,Li2005Structure-PreservingFormulation,Li2005StructuredLMIs}, there is no single iterative method that provides a universally applicable solution to computational limitations. In addition, it is often unclear how the use of such approximate solutions impacts the accuracy of the resulting reduced-order interconnected system model.

Alternatively, in case of sparsely interconnected systems, Leung et al. \cite{Leung2019ModelApproach} perform the structure-preserving reduction of a subsystem using only information on its immediate neighbors. As a result, the reduced subsystem model is relevant with respect to the considered cluster of subsystem models. For each subsystem, its immediate neighbouring subsystems therefore act as a lower-order surrogate for its full environment, consisting of all other subsystem models. By iterating over all subsystems, all reduced subsystem model retain their relevance with respect to their own clusters, theoretically resulting in a (more) accurate reduced, interconnected model.

\begin{figure}
    \subfloat[\small\label{fig:ssRed_RGB}]{%
      \includegraphics[scale = 0.87,valign=t]{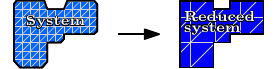}
    }\vspace{-19mm} \\ 
    \subfloat[\small\label{fig:spRed_RGB}]{%
      \includegraphics[scale = 0.87]{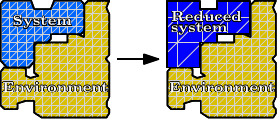}
    }\hfill
    \subfloat[\small\label{fig:absRed_RGB}]{%
      \includegraphics[scale = 0.87]{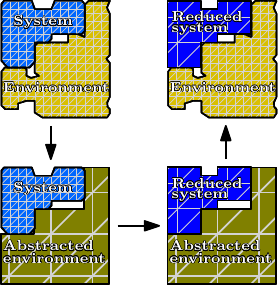}
    }
    \caption{Three modular approaches for the reduction of a system within a dynamic environment: a) Open-loop (independent) reduction of the system model, b) Closed-loop (structure-preserving) reduction of the interconnection of the system and its environment, and c) Abstracted model reduction, where the system is reduced in connection to an abstracted environment model.}
    \label{fig:RGB_red_approaches} 
\end{figure}

In this paper, we introduce a novel, different perspective. Let us focus on a single subsystem, which we then denote as the \emph{system}, which is embedded within an \emph{environment} consisting of the remaining subsystems, as visualized schematically in \autoref{fig:SS-SE_RGB}. A model for this environment is straightforwardly attained by the interconnection of all remaining subsystem models. Using this general system-environment model description, including any system within a dynamic environment, we can schematically visualize the open-loop and closed-loop reduction methods by Figures \ref{fig:ssRed_RGB} and \ref{fig:spRed_RGB}, respectively. Observing the structure-preserving reduction as in \autoref{fig:spRed_RGB}, the environment determines what essential system dynamics to retain (in the light of the interconnected system dynamics). However, this model for the environment is often unnecessarily complex (high order), thereby inducing computation infeasibility. 

\red{To address this challenge, we introduce the framework of \emph{abstracted model reduction}, illustrated in \autoref{fig:absRed_RGB}, as our first main contribution. The core idea is to use a low-order \emph{abstraction} of the environment model, rather than the original, high-order environment model, to identify the system dynamics most relevant to the interconnected system. The framework serves as a versatile solution to improve the efficiency of structure-preserving reduction methods for interconnected systems. }

As a second contribution, we employ techniques from robust performance theory, which have recently been applied in the context of reducing interconnected systems \cite{Janssen2023ModularApproach}, to quantitatively assess how relying solely on an abstraction of the environment influences the resulting accuracy of the reduced interconnected system model. This allows us to establish a priori requirements on the accuracy of both the abstracted environment model and the reduced system model, based on user-defined, frequency-dependent or $\Hnrm{\infty}$-based specifications on the model accuracy of the interconnected system as a whole. These specifications are then leveraged to extend the framework of abstracted model reduction (abstracted reduction for brevity) to the framework of \emph{robust abstracted reduction}. This latter framework automatically generates a reduced-order system model which guarantees stability of the reduced interconnected system model and guarantees the satisfaction of its prescribed accuracy specification.

Finally, we evaluate the abstracted model reduction framework by an industrial case study, where we reduce a structural dynamics model of lithography equipment of the semiconductor industry, consisting of several interconnected subsystem models. Through application of abstracted model reduction in combination with closed-loop balanced reduction \cite{Ceton1993FrequencyReduction,Wortelboer1994Frequency-weightedTools}, we show that a low-order abstraction of a subsystem's environment is sufficient to ensure the relevance of the reduced subsystem model and the accuracy of the reduced interconnected system model. 

The remainder of this paper is organized as follows. In \autoref{sec:prob}, the system-environment representation is presented and the problem is formally defined. Then, the framework of abstracted reduction is introduced in \autoref{sec:absred_framework}. In this framework, an error source is introduced both for the abstraction of the environment and for the reduction of the subsystem. In \autoref{sec:bnd_relations}, we relate both error sources to the resulting error on the level of the interconnected system and use robust performance techniques to relate the different error specifications. This relation is subsequently leveraged to present a systematic approach to abstracted reduction in \autoref{sec:specifications}. Subsequently, in \autoref{sec:case_study_red}, both the general abstracted reduction framework and its robust extension are evaluated and compared to other reduction approaches from literature by means of an industrial case study. Finally, \autoref{sec:con} presents conclusions on the proposed approach.

\emph{Notation:}
In this paper, sets are generally indicated by blackboard-bold symbols, such as $\R$, $\R_{>0}$ and $\C$, which denote the set of real, positive real and complex numbers, respectively. $\R^{m\times p}$ and $\C^{m\times p}$ indicate matrices of real and complex numbers, respectively, with $m$ rows and $p$ columns. Given a complex matrix $A$, $A^\top$ and $A^H$ denote its transpose and conjugate transpose, respectively, $\|A\|$ denotes its 2-induced norm, $A\succ 0$ and $A\succeq 0$ denote that $A$ is positive definite and positive semi-definite, respectively, and $A = \diag(A_1,A_2)$ denotes a block-diagonal matrix of submatrices $A_1$ and $A_2$. The zero matrix and identity matrix are denoted by $O$ and $I$, respectively, while $I_n$ denotes an identity matrix of size $n$. Given a transfer function matrix $\sys(s)$, where $s$ is the Laplace variable, $\|\sys\|_\infty$ denotes its $\Hnrm{\infty}$-norm. The set of all proper, real rational stable transfer matrices is denoted by $\RH$.

%
%%%%%%%%%%%%%%%%%%%%%%%%%%%%%%%%%%%%%%%%%
%
\section{Problem setting}\label{sec:prob}

\subsection{Interconnected system representation}\label{ssec:sys_repr}
\begin{figure}
    \subfloat[\small\label{fig:coupling_diag}]{%
      \includegraphics[width=0.45 \linewidth]{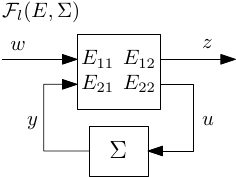}
    }
    \hfill
    \subfloat[\small \label{fig:coupling_diag_red}]{%
      \includegraphics[width=0.45\linewidth]{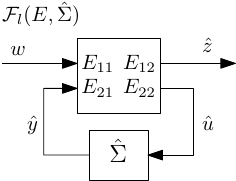}
    }
    \caption{(a) Lower LFT of $\env(s)$ and $\sys(s)$, constituting the interconnected model $\llft(\env,\sys)$ and (b) lower LFT of $\env(s)$ and $\sysr(s)$, constituting the reduced, interconnected model $\llft(\env,\sysr)$.}
    \label{fig:coupling_diags}\vspace{-4mm}
\end{figure}

% In Sections \ref{sec:prob}-\ref{sec:specifications}, the subsystem to be reduced will be referred to as the \emph{system}. The interconnection of the remaining subsystems still will be referred to as the \emph{environment} and the interconnection of all subsystems still will be referred to as the \emph{interconnected system}. 
The system and its environment, as schematically visualized in \autoref{fig:SE_RGB}, are modeled by the proper, real rational transfer function matrices $\sys(s)$ and $\env(s)$, respectively, such that the interconnected system model can be described by the block-diagram shown in \autoref{fig:coupling_diag}. The system model $\sys(s)$ has inputs $u\in \R^m$, outputs $y\in \R^p$ and McMillan degree (order) $n_\sys$ and the environment model $\env(s)$ has inputs $w\in \R^{m_C}$ and $y\in \R^p$, outputs $z\in\R^{p_C}$ and $u\in \R^m$ and order $n_\env$, such that
\begin{equation} \label{eq:env_partition}
    \begin{bmatrix}
        z \\ u
    \end{bmatrix} = 
    \begin{bmatrix}
        E_{11} & E_{12}\\ E_{21} & E_{22}
    \end{bmatrix}
    \begin{bmatrix}
        w \\ y
    \end{bmatrix}.
\end{equation}
The system and environment are interconnected by means of a lower linear fractional transformation (LFT), as defined below.

\begin{defn} \label{def:llft_ulft_wellposed}
    Let $P(s) = \left[\begin{smallmatrix}  P_{11}(s) & P_{12}(s)\\ P_{21}(s) & P_{22}(s) \end{smallmatrix}\right]$, $M_l(s)$ and $M_u(s)$ be proper, real rational transfer function matrices of dimensions $(p_1+p_2)\times(m_1+m_2)$, $m_2\times p_2$ and $m_1\times p_1$, respectively. Then, we define the lower and upper LFTs, respectively, as 
    \begin{align}
        \llft(P,M_l) &= P_{12}M_l(I-P_{22}M_l)^{-1}P_{21}+P_{11}, \label{eq:llft_def} \\ 
        \ulft(P,M_u) &= P_{21}M_u(I-P_{11}M_u)^{-1}P_{12}+P_{22},\label{eq:ulft_def}
    \end{align}    
    which are said to be \emph{well-posed} if $I-P_{22}M_l$ and $I-P_{11}M_u$ have a proper real rational inverse, respectively \cite[Def.~9.2,~Lem.~5.1]{Zhou1998EssentialsControl}.
\end{defn}

Using this definition, the interconnection of system $\sys$ and environment $\env$ as in \autoref{fig:coupling_diag} can be written as $\llft(\env,\sys)$. We make the following assumption to ensure that $\llft(\env,\sys)$ is well-defined and internally stable (see \citep[Chapter~5]{Zhou1998EssentialsControl}).

\begin{ass}\label{ass1} 
    We have $\sys(s)\in\RH$, $\env(s)\in\RH$ and the interconnection $\llft(\env,\sys)$ is well-posed and internally stable, i.e., $\sys(I-\env_{22}\sys)^{-1} \in \RH$. Particularly, we have $\llft(\env,\sys) \in \RH$.
\end{ass}
In the setting of interconnected systems, $\env(s)$ typically represents several (interconnected) subsystems such that its order is typically large, particularly $n_E>n_\sys$. This motivates the current work, where the evaluation of $\sys(s)$ is computationally tractable, but the evaluation of $\llft(\env,\sys)$ is not.

\vspace{-1.5mm}
\subsection{Structure-preserving model reduction} \label{ssec:structpres} \vspace{-0.5mm}
The high-level objective is to find a transfer function matrix $\sysr(s)$ of order $r_\sys < n_\sys$ such that $\llft(\env,\sysr)$, as depicted in \autoref{fig:coupling_diag_red}, is well-posed, stable and accurately approximates $\llft(\env,\sys)$ in terms of input-output behaviour, i.e., such that the approximation error, being the output of the error dynamics:
\vspace*{-.75mm}
\begin{equation} \label{eq:ec_init_def}
    \ec \coloneqq \llft(\env,\sysr)- \llft(\env,\sys),  
\vspace*{-.75mm}
\end{equation}
is small in a suitable sense. Structure-preserving reduction methods as discussed in \cite{Sandberg2009,Vandendorpe2008ModelSystems,Poort2024Balancing-BasedSystems,Cheng2019BalancedSystems} aim to find an accurate reduced-order model $\llft(\env,\sysr)$. However, they are often only applicable to interconnected systems of limited order. In addition, most methods either do not guarantee stability of $\llft(\env,\sysr)$, e.g. \cite{Sandberg2009,Vandendorpe2008ModelSystems}, or require additional system properties such as passivity \cite{Poort2024Balancing-BasedSystems,Cheng2019BalancedSystems}.

A further, intrinsic limitation of structure-preserving reduction methods is the need for environment model $\env(s)$. In a modular, model-based design process of complex engineering systems, where subsystems are designed in parallel, the environment model $\env(s)$ is typically not available or only a rough estimate $\envr(s)$ is available.

\vspace{-1.5mm}
\subsection{Problem statement}\label{ssec:prob_stat} \vspace{-0.5mm}
Our goal is to address the limitations of existing structure-preserving model reduction methods. Particularly, given a system $\sys(s)$ and environment $\env(s)$, possibly both of large order, we aim to reduce $\sys(s)$ to $\sysr(s)$ such that
\begin{enumerate}
    \item stability is preserved, i.e., $\llft(\env,\sysr)$ is well-posed, internally stable, and $\llft(\env,\sysr) \in \RH$,
    \item the approximation is accurate in the sense that the approximation error dynamics $\ec \coloneqq \llft(\env,\sysr)- \llft(\env,\sys)$ is small.
\end{enumerate} 

Specifically, we consider the case where the order of $\env(s)$ is high, such that the application of existing structure-preserving reduction methods to $\llft(\env,\sys)$ is not feasible (or comes at too large computational cost).

%
%%%%%%%%%%%%%%%%%%%%%%%%%%%%%%%%%%%%%%%%%
%
\vspace{-1mm}
\section{Abstracted reduction framework}\label{sec:absred_framework}
We will initially neglect giving guarantees on the accuracy and stability of the reduced model (such guarantees will be treated in Sections \ref{ssec:robperf} and \ref{sec:specifications}) and focus on addressing the computational limitations of structure-preserving reduction methods. To this end, we propose the abstracted reduction framework in \autoref{ssec:absred-alg}. The computational benefits of this approach and some applicability considerations are subsequently discussed in \autoref{ssec:absred-appl}.

\vspace{-1.5mm}
\subsection{The abstracted reduction algorithm} \label{ssec:absred-alg}
\begin{figure}
    \centering
    \includegraphics[width = \linewidth]{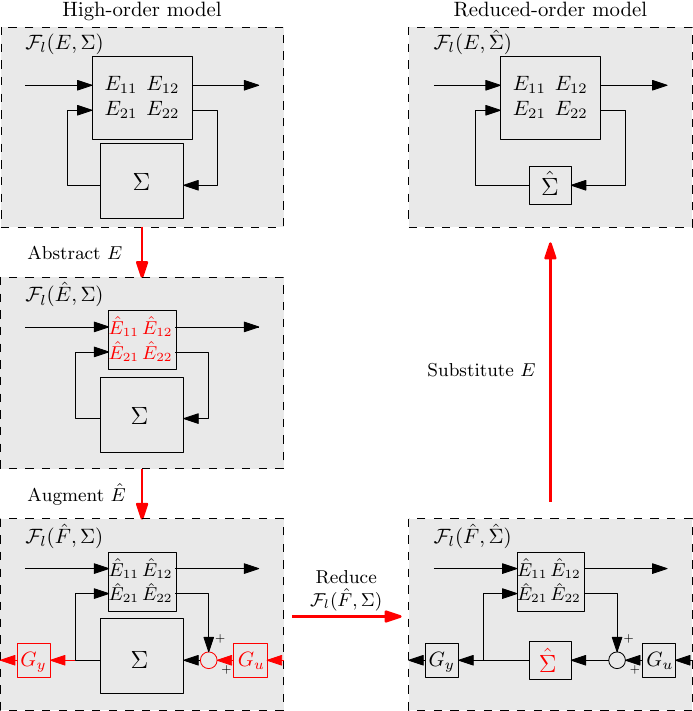}
    \caption{Schematic representation of the steps of abstracted reduction.}
    \label{fig:absred_steps} \vspace{-2mm}
\end{figure}

To facilitate efficient and accurate model reduction, we present our framework of abstracted reduction, which is also illustrated in \autoref{fig:absred_steps}, consisting of the following steps.
\vspace{-2mm}

\begin{alg} \label{alg:absred}
    \emph{Abstracted reduction}\\
    \textbf{Input:} $p\times m$ transfer matrix $\sys(s)$ and $(p_C+m)\times(m_C+p)$ transfer matrix $\env(s)$, of orders $n_\sys,\ n_\env$, respectively, abstraction order $r_E\leq n_E$ and reduction order $r_\sys<n_\sys$ and weighting matrices $G_y\in\C^{p\times p}$ and $G_u\in\C^{m\times m}$.\\
    \textbf{Output:} Surrogate model $\sysr(s)$ of reduced order $r_\sys$, such that $\llft(\env,\sysr)$ approximates $\llft(\env,\sys)$.
\begin{enumerate}
    \item \textbf{Abstraction of $\env(s)$}. Abstract $\env(s)$ to $\envr(s)$ in open loop, i.e., disconnected from $\sys(s)$, by means of, e.g., reduction or another form of surrogate modelling.
    \item \textbf{Augmentation of $\envr(s)$}. Augment the set of external inputs and outputs of $\llft(\envr,\sys)$ by incorporating $\sys$'s (weighted) inputs $u$ and outputs $y$. This is equivalent to replacing $\envr(s)$ with $\fnvr(s)$, resulting in $\llft(\fnvr,\sys)$, where
    \begin{equation} \label{eq:F_defs}
        \fnvr(s) = \begin{bmatrix}    \fnvr_{11} & \fnvr_{12}\\ \fnvr_{21} & \fnvr_{22}   \end{bmatrix} 
        = \left[\begin{array}{cc|c} \envr_{11} & O & \envr_{12}\\ O &O &G_y\\ \hline \envr_{21} & G_u & \envr_{22}\end{array}\right],
    \end{equation}
    where $\envr$ is partitioned the same as $\env$, see \autoref{eq:env_partition}.
    \item \textbf{Reduction of $\sys(s)$}. Use a structure-preserving reduction method to reduce $\llft(\fnvr,\sys)$ to $\llft(\fnvr,\sysr)$.
    \item \textbf{Substitution of $\env(s)$}. Substitute the original environment $\env(s)$ to obtain $\llft(\env,\sysr)$.
\end{enumerate}
\end{alg}

To interpret the concept of abstracted reduction, let us first \red{disregard the augmentation step by selecting} $G_u(s) = O_{m\times m}$, $G_y(s) = O_{p\times p}$ and assume the use of reduction methods that approximate a model's input-output behaviour, such as balancing methods \cite{Wortelboer1994Frequency-weightedTools,Moore1981PrincipalReduction}. Then, abstracted reduction can be interpreted as follows: whereas an open-loop reduction approach evaluates the input-output behaviour of the system ($\sys$) itself and structure-preserving reduction approach evaluates the \emph{coupled} input-output behaviour (of $\llft(\env,\sys)$), the \emph{abstracted} reduction approach uses structure-preserving reduction methods to evaluate the \emph{approximate, coupled} input-output behaviour (via $\llft(\fnvr,\sys)$), thereby significantly reducing computational costs. In an ideal world, without computational limitations, one would like to do direct, structure-preserving reduction of $\llft(\env,\sys)$. However, this is not always feasible or it is computationally very expensive. Hence, we suggest abstracted reduction as a framework to enable an approximate structure-preserving reduction, where the direct structure-preserving reduction of $\llft(\env,\sys)$ is not feasible or too computationally expensive.

\red{The augmentation with the weighting matrices $G_u$ and $G_y$ in \autoref{alg:absred} serves two main purposes: (i) to control the accuracy distribution over $\sysr$ and $\llft(E,\sysr)$, and (ii) to facilitate error analysis in \autoref{sec:bnd_relations}. Regarding the first purpose, increasing the magnitudes of $G_u$ and $G_y$ amplifies the contribution of $\sys$'s input-output behaviour in $\llft(\fnvr,\sys)$. When using an input-output approximating reduction method, high-magnitude $G_u$ and $G_y$ thus improve $\sysr$'s approximation of $\sys$ at the cost of the accuracy of $\llft(\fnvr,\sysr)$. This trade-off is discussed further in \autoref{ssec:considerations}. Regarding the second-purpose, it turns out that extending the number inputs and outputs $\llft(\envr,\sys)$ to $\llft(\fnvr,\sys)$, by augmentation of $\envr(s)$ to $\fnvr(s)$, is often essential for deriving the error relations in the abstracted reduction algorithm, as detailed in \autoref{sec:bnd_relations}.}

\subsection{Computational benefits and applicability} \label{ssec:absred-appl}
To compare the computational cost of abstracted reduction to direct structure-preserving reduction of $\llft(\env,\sys)$, recall that the model orders of $\env(s)$, $\envr(s)$, $\sys(s)$ and $\sysr(s)$ are denoted as $n_E$, $r_E$, $n_\sys$ and $r_\sys$, respectively. Structure-preserving reduction methods based on balancing, such as \cite{Vandendorpe2008ModelSystems,Poort2024Balancing-BasedSystems}, usually scale cubically with the order of $\llft(\env,\sys)$ \cite{Antoulas2005AnSystems}, while structure-preserving methods based on LMI's, such as \cite{Li2005StructuredLMIs,Beck1996ModelSystems}, scale even more steeply. Hence, our abstracted approach replaces a cost $(n_\env\!+\!n_\sys)^c$ with $(r_E\!+\!n_\Sigma)^c$ for $c\! \geq\! 3$, i.e., the en-vironment is replaced by a low-order abstraction, which yields a significant reduction in computational cost when $n_\env$ is large and $r_\env$ is small, and when $n_\env$ is large compared to $n_\sys$.

\begin{figure}
    \centering
    \includegraphics[width = 0.7\linewidth]{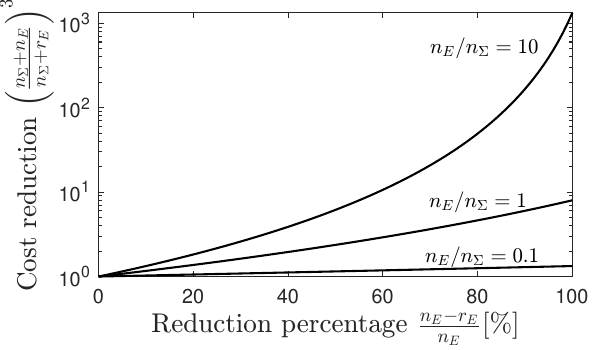}
    \caption{Computational cost reduction per the amount of reduction of $\env(s)$, assuming negligible abstraction cost and cubic scaling of the reduction's cost. The results are given for three different relative orders of $\env(s)$ and $\sys(s)$.}
    \label{fig:cost reduction}
\end{figure}

The low-order abstraction $\envr$ can be obtained by an inexpensive reduction method, which does not preserve structure, such as \cite{Gugercin2008H2Systems,Craig1981Component-ModeSynthesis}, as its accuracy is not so important for the accuracy of the reduced, interconnected system $\llft(\env,\sysr)$; it merely acts as a weight in the reduction of $\sys$ to $\sysr$. In such a case, the computational cost of the reduction of $\sys$ to $\sysr$ is dominant. This makes abstracted reduction mostly beneficial to systems where $n_E>n_\sys$ and when $r_E$ is selected as $r_E\,\red{\ll}\, n_E$, as shown in \autoref{fig:cost reduction}.

% \begin{exmp}\label{ex:comp_cost}
%     Consider transfer function matrices $\sys(s)$ and $\env(s)$ of orders 1000 and 10 000, respectively. If abstracted reduction is performed using an abstraction $\envr(s)$ of order 100 ($99\%$ reduction), the computational cost is approximately $10^{11}$ flops, while direct reduction requires $10^{14}$ flops.
% \end{exmp}

\begin{rem} 
    Sometimes, $\sys(s)$ needs to be reduced while having incomplete knowledge of $\env(s)$. This is a common occurrence in modular design processes, where several subsystems are designed simultaneously. However, these design processes are typically iterative, such that we generally possess subsystem models from a previous iteration. In our abstracted approach, we suggest to use $\envr(s)$, an approximation of $\env(s)$, to effectively indicate which dynamics of $\sys(s)$ are relevant to $\llft(\env,\sys)$. Therefore, an $\envr(s)$ of a previous design iteration or even an $\envr(s)$ based on a basic, preliminary environment model might improve the accuracy of $\llft(\env,\sysr)$ beyond what can be achieved through (open-loop) reduction of $\sys$. However, in case a preliminary model $\envr$ is used without knowledge of its accuracy (i.e., without knowing $\env$), the formal error analysis of \autoref{sec:bnd_relations} can not be applied.
\end{rem}

%
%%%%%%%%%%%%%%%%%%%%%%%%%%%%%%%%%%%%%%%%%
%
\section{Error analysis and the relation of bounds} \label{sec:bnd_relations}
The reduction goal, as specified in \autoref{sec:prob}, is to achieve a low-order surrogate model $\sysr(s)$ such that the reduced, interconnected model $\llft(\env,\sysr)$ is well-posed, stable and (accurately) approximates $\llft(\env,\sys)$.
To this end, we first assume the reduction/abstraction errors to be known transfer functions and determine the relation between introduced reduction and abstraction errors and the error of the interconnected system in \autoref{ssec:error_charac}. Subsequently, building on the approach introduced in \cite{Janssen2024ModularPerspective}, we relate bounds on these errors in \autoref{ssec:robperf}.

%%%%%%%%%%%%%%%%%%%%%%%%%%%%%%%%%
\subsection{Error relations}\label{ssec:error_charac}
We start by defining the errors introduced by the abstracted reduction procedure described by \autoref{alg:absred} in \autoref{ssec:absred-alg}. First, the abstraction of the environment $\env(s)$ to its low-order approximation $\envr(s)$ leads to the error system
\begin{equation}\label{eq:abs_error_def}
         \ee(s) \coloneqq\envr(s) - \env(s).
\end{equation}
We recall from step 2 of \autoref{alg:absred} that the resulting system $\llft(\envr,\sys)$ is subsequently augmented to obtain the system $\llft(\fnvr,\sys)$. Augmentation does not affect the well-posedness and stability of the interconnection, as stated next.
\begin{lem}
    Consider the $p\times m$ transfer matrix $\sys(s)$ and $(p_C+m)\times(m_C+p)$ transfer matrix $\envr(s)$, such that $\llft(\envr,\sys)$ is well-posed and internally stable. Then, for any $\fnvr(s)$ in \autoref{eq:F_defs}, with weighting matrices $G_y\in\C^{p\times p}$ and $G_u\in\C^{m\times m}$, $\llft(\fnvr,\sys)$ is well-posed and internally stable.
\end{lem}
\begin{proof}
    Following \autoref{def:llft_ulft_wellposed} and \cite[Corollary~5.2]{Zhou1998EssentialsControl},  $\llft(\fnvr,\sys)$ is well-posed and internally stable if $I-\envr_{22}\sys$ has a proper real rational inverse and $\sys(I-\envr_{22}\sys)^{-1}\in \RH$, respectively. Both notions follow directly from the well-posedness and internal stability of $\llft(\envr,\sys)$. 
\end{proof}
Next, the application of structure-preserving reduction methods to $\llft(\fnvr,\sys)$ (reduction step at the bottom of \autoref{fig:absred_steps}) leads to a reduced-order system $\sysr$ and error system
\begin{equation}\label{eq:red_error_def}
         \ef(s) \coloneqq\llft\big(\fnvr(s), \sysr(s)\big) - \llft\big(\fnvr(s), \sys(s)\big).
\end{equation}
We emphasize that we work with the error system $\ef(s)$ rather than with $\sysr(s)-\sys(s)$, for the following two reasons:
\begin{itemize}
    \item Structure-preserving reduction methods reduce subsystems based on the coupled dynamics, such that any available error bounds are typically bounds on $\ef(s)$.
    \item As we aim to (accurately) approximate the \emph{coupled} dynamics, the magnitude of $\sysr-\sys$ cannot be expected to be a good indication of the quality of reduction. Alternatively, $\ef$ typically gives a much better indication of the quality of the overall reduction, as long as $\ee$ is small. Note that if $\ee = 0$, $G_u = O_{m\times m}$ and $G_y = O_{p\times p}$, indeed $\ef = \llft(\env,\sysr)- \llft(\env,\sys)$. Therefore, if a specification on the accuracy of $\llft(\env,\sysr)$ is translated to bounds on $\ee$ and $\ef$, these bounds are expected to be less conservative than similar bounds on $\sysr-\sys$.
\end{itemize}

The abstraction and reduction steps, characterized through the errors in \autoref{eq:abs_error_def} and \autoref{eq:red_error_def}, respectively, ultimately lead to the reduction error $\ec$ of the interconnected system, as previously introduced in \autoref{eq:ec_init_def} and repeated here for completeness as
\begin{equation}\label{eq:cpld_error_def}
    \ec(s) \coloneqq \llft\big(\env(s),\sysr(s)\big)- \llft\big(\env(s),\sys(s)\big).
\end{equation}

Our main goal in this section is to relate $\ec$ to the error systems $\ee$ and $\ef$. As a first step in this direction, we express the reduced-order system $\sysr$ in terms of the reduction error system $\ef$.

\begin{lem}\label{lem:error_corr}
    Let $\sys(s)$, $\sysr(s)$ be transfer function matrices and let $\fnvr(s)$ be as in \autoref{eq:F_defs} with square, invertible matrices $G_u$ and $G_y$, such that $\llft(\fnvr,\sys)$ and $\llft(\fnvr,\sysr)$ are well-posed with a difference $\ef(s)$ as in \autoref{eq:red_error_def}. 
    Then
    \begin{equation}\label{eq:EClem_sysr}
        \hspace{-1mm} \sysr = \ulft\!\left(\left[\begin{smallmatrix}\!-\envr_{22}&I\\\!I&O\end{smallmatrix}\right], \sys(I-\envr_{22}\sys)^{-1} \!+ G_y^{-1} \efs{22} G_u^{-1} \right),
    \end{equation} 
    where 
    \begin{equation} \label{eq:ef22_Def}
        \efs{22} = \llft\Big(\left[\begin{smallmatrix}  O&G_y\\G_u & \envr_{22} \end{smallmatrix}\right], \sysr\Big) - \llft\Big(\left[\begin{smallmatrix}  O&G_y\\G_u & \envr_{22}\end{smallmatrix}\right], \sys\Big) 
    \end{equation}
    is the $22$-partition of $\ef$ as in \autoref{eq:red_error_def}, representing the transfer matrix from the augmented inputs to the augmented outputs of $\llft(\fnvr,\sys)$.
\end{lem}
\begin{proof}
    Let us first define $P\coloneqq \left[\begin{smallmatrix}  O&G_y\\G_u & \envr_{22} \end{smallmatrix}\right]$ and note that \autoref{eq:ef22_Def} can be rewritten as 
    \begin{equation} \label{eq:ef22_Pdef}
        \llft\big(P,\sysr\big) = G_y\sys(I-\envr_{22}\sys)^{-1}G_u+ \efs{22}.
    \end{equation}
     Given the fact that $G_u$ and $G_y$ are square and invertible, we can then use the inversion formula of \cite[Lemma~10.4]{Zhou1996RobustControl} to extract $\sysr$ from $\llft(P,\sysr)$ as
    \begin{equation}\label{eq:EClem_sysr_short}
        \sysr = \ulft\left(P^{-1},\llft(P,\sysr)\right),
    \end{equation}
    where 
    \begin{equation}\label{eq:Pinv}
    P^{-1} = \left[\begin{smallmatrix}-G_u^{-1}\envr_{22}G_y^{-1}&G_u^{-1}\\G_y^{-1}&O\end{smallmatrix}\right].  
    \end{equation}
    A substitution of \autoref{eq:ef22_Pdef} and \autoref{eq:Pinv} into \autoref{eq:EClem_sysr_short} gives an expression similar to \autoref{eq:EClem_sysr}. Using the definition of the upper LFT in \autoref{eq:ulft_def}, we finally move the $G_u^{-1}$ and $G_y^{-1}$ terms from $P^{-1}$ to $\llft\big(P,\sysr\big)$ to attain the expression for $\sysr$ as in \autoref{eq:EClem_sysr}.

\end{proof}

\begin{figure}
    \centering
    \includegraphics{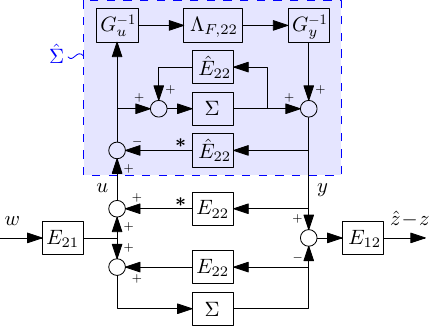}
    \caption{Schematic representation of the coupled error system $\ec(s)\coloneqq\llft(\env,\sysr)-\llft(\env,\sys)$, where $\sysr$ is given as in \autoref{eq:EClem_sysr}.}
    \label{fig:err_sys}
\end{figure}

\begin{rem}
    The inversion formula of \cite[Lemma~10.4]{Zhou1996RobustControl} used in the proof of \autoref{lem:error_corr} motivates the augmentation of $\env$ with $G_u$ and $G_y$. Note, however, that we used \emph{only} the augmented inputs and outputs to express $\sysr$. This augmentation is excessive as we can also express $\sysr$ in terms of $\ef$ instead of $\efs{22}$. When using $\ef$, $P=\fnvr$ in the proof, and the inversion requires $\fnvr_{21}$ and $\fnvr_{12}$ to be full column and row rank, respectively. In this general case, $\sysr$ can be expressed by \autoref{eq:EClem_sysr}, with $G_y^{-1}\efs{22}G_u^{-1}$ replaced by $\fnvr_{12}^\dagger \ef \fnvr_{21}^\dagger$, where $(.)^\dagger$ denotes the Moore-Penrose pseudoinverse. Augmentation of $\env$ with square and invertible $G_u$ and $G_y$, however, gives the most convenient expression for $\sysr$, because this prevents the inversion of (unknown) transfer function matrices $\fnvr_{21}$ and $\fnvr_{12}$.
\end{rem}
    
Using the lower LFT definition in \autoref{eq:llft_def}, the expression for the total reduction error system $\ec$, as in \autoref{eq:cpld_error_def}, can be rewritten to
\begin{equation}
    \ec = \env_{12}\left( \sysr(I-\env_{22}\sysr)^{-1} - \sys(I-\env_{22}\sys)^{-1}  \right)\env_{21}.
\end{equation}
Substitution of $\sysr(s)$ as in \autoref{eq:EClem_sysr} gives an expression of $\ec$ in terms of the error system $\efs{22}$, as depicted schematically in \autoref{fig:err_sys}, where the expression for $\sysr(s)$ is highlighted. This block diagram facilitates an intuitive interpretation. Firstly, if $\efs{22} = 0$, the two feedback-loops with $\envr(s)$ cancel and $\sysr(s) = \sys(s)$. This represents the inverse relation between the abstraction and substitution steps of the abstracted reduction approach. Also, if $\envr_{22}(s) = \env_{22}(s)$, the feedback-loops indicated by a $\bm{\ast}$ in \autoref{fig:err_sys} cancel and the approach simplifies to standard structure-preserving reduction.

The observed dependency of $\ec$ on $\envr_{22}$ in \autoref{fig:err_sys} suggests that $\envr_{11}$, $\envr_{12}$ and $\envr_{21}$ do not influence $\ec$. Note, however, that $\envr$ determines $\llft(\fnvr,\sys)$ through the definition of $\fnvr$ in \autoref{eq:F_defs}. $\efs{22}$ thus implicitly depends on $\envr$. We will not consider this implicit dependency in the remainder of this paper and treat $\efs{22}$ as a independent error source.

Motivated by \autoref{fig:err_sys}, we obtain a characterization of $\ec$ that allows us to isolate the influence of abstraction and reduction errors on the error $\ec$ on the reduced-order interconnected system.
\begin{thm}\label{thm:ec_lft}
    Let $\sys(s)$, $\sysr(s)$ be transfer function matrices and let $\fnvr(s)$ be as in \autoref{eq:F_defs}, satisfying \autoref{ass1}. Let $\ec(s)$ and $\efs{22}(s)$ be as in \autoref{eq:cpld_error_def} and \autoref{eq:ef22_Def}, respectively, and let $\ees{22} \coloneqq \envr_{22} - \env_{22}$ and $\eft \coloneqq G_y^{-1} \efs{22} G_u^{-1}$, then
    \begin{equation}\label{eq:ec_lft}
            \ec = \ulft(N,\diag(\ees{22},\eft,\ees{22}))
    \end{equation}
    where
    \begin{equation} \label{eq:N_def}
        N \coloneqq\left[\begin{array}{ccc|c}
            M & I & M & M\env_{21}\\
            I & O & O & \env_{21}\\
            M & O & M & M \env_{21}\\ \hline
            \env_{12}M & \env_{12} & \env_{12}M & O\\
        \end{array}\right], 
    \end{equation}
    and
    \begin{equation} \label{eq:N_def_M}
            M(s)= \sys(s)(I-\env_{22}(s)\sys(s))^{-1},
    \end{equation} 
    such that $N(s)\in \RH$ due to \autoref{ass1}.
\end{thm}
\begin{proof}
    Substitution of $\envr_{22} = \env_{22} + \ees{22}$, $\eft(s) \coloneqq G_y^{-1} \efs{22}(s) G_u^{-1}$ and \autoref{eq:EClem_sysr} into $\ec \coloneqq \llft(\env,\sysr)-\llft(\env,\sys)$ gives
    \begin{multline}\label{eq:ec_expr}
        \ec =  -\llft(\env,\sys) + \llft\left(\env,\ulft\left(\left[\begin{smallmatrix}-\env_{22}-\ees{22}&I\\I&O\end{smallmatrix}\right],\,\right.\right.\\
        \left.\left. \sys(I-(\env_{22}+\ees{22})\sys)^{-1} + \eft \right)\right).
    \end{multline} 
    To extract the error terms of $\ees{22}$ and $\eft$, we note that 
    \begin{equation}
        \begin{aligned}
            \ulft\big(\left[\begin{smallmatrix} \env_{22}+\ees{22}&I\\I&O \end{smallmatrix}\right],\sys\big) &= \sys(I-(\env_{22}+\ees{22})\sys)^{-1},\\
            &=\ulft\big(\left[\begin{smallmatrix} M&M\\M&M \end{smallmatrix}\right],\ees{22}\big),
        \end{aligned}
    \end{equation}
    with $M$ as in \autoref{eq:N_def_M}, such that \autoref{eq:ec_expr} can be rewritten as
    \begin{equation}
        \ec = -\llft(\env,\sys) + \llft\left(\env,
        \ulft\left(\left[\begin{smallmatrix}K&K\\K&K\end{smallmatrix}\right],-\ees{22}\right)\right),
    \end{equation}
    where
    \begin{equation}
        K = \ulft\left(\left[\begin{smallmatrix} -\env_{22}&I\\I&O \end{smallmatrix}\right],\ulft\left(\left[\begin{smallmatrix}M&M\\M&M\end{smallmatrix}\right],\ees{22}\right) + \eft \right).
    \end{equation}
    Finally, these nested LFT's can be expressed as the LFT of \autoref{eq:ec_lft}, making repeated use of \cite[Lemma~10.3]{Zhou1996RobustControl}.
\end{proof}

\begin{figure}
    \centering
    \includegraphics[width= \linewidth]{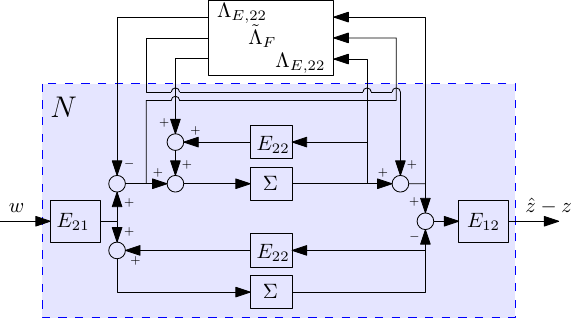}
    \caption{The coupled error dynamics $\ec(s)$ of \autoref{eq:ec_lft}, expressed as an upper LFT of the nominal model $N(s)$ and error terms $\ees{22}$ and $\eft$.}
    \label{fig:ec_lft}
\end{figure}

The expression of $\ec$ as in \autoref{eq:ec_lft} can be verified intuitively by its graphical representation in \autoref{fig:ec_lft}. Starting from \autoref{fig:err_sys}, which represents the definition of $\ec$ in \autoref{eq:cpld_error_def}, we express $\envr_{22}$ as a parallel connection of $\env_{22}$ and $\ees{22}$ and ``pull out'' the error terms $\ees{22}$ and $\eft$, resulting in \autoref{fig:ec_lft}. This shows that, indeed, \autoref{eq:ec_lft} expresses $\ec$ as in \autoref{eq:cpld_error_def}.

%%%%%%%%%%%%%%%%%%%%%%%%%%%%%%%%%
\subsection{A robust performance perspective}\label{ssec:robperf}
Next, we will use the formulation of the overall error dynamics as in \red{\autoref{thm:ec_lft}} (and \autoref{fig:ec_lft}) to relate requirements on bounds on this error dynamics to the errors made in the underlying abstraction and (structure-preserving) reduction step. We will assume a prescribed, weighted bound on the error $\ec(s)$ and aim to translate this to bounds on $\ees{22}(s)$ and $\eft(s)$. Specifically, we prescribe weighting matrices $V_C(s)\in \RH$ and $W_C(s) \in \RH$ and formulate the requirement as
\begin{equation}\label{eq:dc}
    \dc(s) = V_C(s)\, \ec(s)\, W_C(s), 
\end{equation}
for some $\dc(s) \in \RH$ satisfying \red{$\|\dc\|_\infty \leq 1$}.

The abstraction and reduction errors $\ees{22}(s)$, $\eft(s)$ are expressed similarly as
\begin{equation}\label{eq:des_dft}
    \begin{aligned}
        \ees{22}(s) &= W_\env(s) \,\des{22}(s) \,V_\env(s),\\
        \eft(s) &= W_F(s) \,\dft(s) \,V_F(s),
    \end{aligned}
\end{equation}
for \emph{bistable} weighting matrices $V_\env,\,W_\env,\,V_F,\,W_F\in \RH$, implying that $V_\env^{-1},\,W_\env^{-1},\,V_F^{-1},\,W_F^{-1}\in \RH$, and some $\des{22},\, \dft\in \RH$ satisfying 
\begin{equation}
    \|\des{22}\|_\infty \leq 1, \quad \|\dft\|_\infty \leq 1.
\end{equation}

The requirements of \autoref{eq:dc} and \autoref{eq:des_dft} can be stated equivalently by restricting $\ees{22},\,\eft$ and $\ec$ to the sets of transfer functions $\uees{22}$, $\ueft$ and $\uec$, respectively, given as
\begin{equation}\label{eq:acc_specs}
    \begin{aligned}
        \uees{22} & \coloneqq\big\{\ees{22}\ \big|\  \|W_E^{-1} \ees{22} V_E^{-1}\|_\infty \leq 1 \big\}, \\ 
        \ueft & \coloneqq\big\{\eft\ 
        \big|\  \|W_F^{-1} \eft V_F^{-1}\|_\infty \leq 1 \big\}, \\
        \uec & \coloneqq\big\{\ec\ \big| \ \|V_C \ec W_C\|_\infty < 1 \big\}.
    \end{aligned}
\end{equation}

% We can now formulate the goal of this section explicitly as: given a specification $\uec$, determine the requirements $\uees{22}$ and $\ueft$, such that for any $\ees{22}\in\uees{22}$ and $\eft\in\ueft$, the coupled error dynamics is guaranteed to satisfy $\ec\in\uec$.
We can now formulate the goal of this section explicitly as: \red{given the sets $\bbLambda_{E,22}$, $\tilde{\bbLambda}_F$ and $\bbLambda_C$, determine whether the coupled error dynamics are guaranteed to satisfy $\ec\in\bbLambda_C$ for any $\ees{22}\in\bbLambda_{E,22}$ and $\eft\in\tilde{\bbLambda}_F$.}

\begin{figure}
    \centering
    \includegraphics[width= 0.5\linewidth]{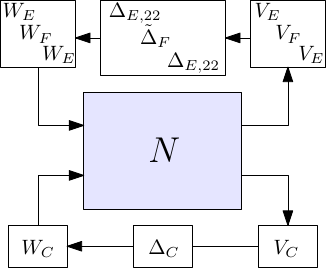}
    \caption{Robust performance framework, with the nominal model $N$ and the weighted uncertainties pulled out.}
    \label{fig:rob_perf_diagram}
\end{figure}

\red{To determine the relation between $\bbLambda_{E,22}$, $\tilde{\bbLambda}_F$ and $\bbLambda_C$}, we will use tools from robust performance analysis. By combining expression \autoref{eq:ec_lft} for the coupled error dynamics $\ec$ with \autoref{eq:dc} and \autoref{eq:des_dft}, we attain the framework of robust performance theory, as visualized in \autoref{fig:rob_perf_diagram}. Then, the uncertainties and weights are combined in the block-diagonal transfer functions
\begin{equation}\label{eq:DVW_tfs}
    \begin{aligned}
        \Delta(s) &\coloneqq \diag(\des{22}(s),\dft(s),\des{22}(s),\dc(s)),\\
        V(s) &\coloneqq \diag(V_\env(s),V_F(s),V_\env(s),V_C(s)),\\
        W(s) &\coloneqq \diag(W_\env(s),W_F(s),W_\env(s),W_C(s)).
    \end{aligned}
\end{equation} 

Furthermore, we introduce scaling matrices $D_\ell$ and $D_r$, which satisfy $D_r^{1/2} \Delta = \Delta D_\ell^{1/2}$ for any $\Delta$ of the form \autoref{eq:DVW_tfs} \cite[Theorem~3.8]{Packard1993TheValue}, by requiring $(D_\ell,D_r)\in\mathbb{D}$, with 
\begin{align}\nonumber
            \mathbb{D} \coloneqq \bigg\{ (D_\ell,D_r) \,\bigg|\, D_\ell &= \scalebox{0.9}{$\begin{bmatrix}
                S_{11}I_{p}& O& S_{12}I_{p}& O\\
                O& d_FI_{m}&O& O\\
                S_{21}I_{p}& O &S_{22}I_{p}&O \\
                O&O&O& I_{m_C}\\
            \end{bmatrix}$}, \\ \label{eq:D_def}
            D_r &= \scalebox{0.9}{$\begin{bmatrix}
                S_{11}I_{m}& O&S_{12}I_{m}& O\\
                O& d_FI_{p}&O& O\\
                S_{21}I_{m}& O &S_{22}I_{m}&O \\
                O&O&O& I_{p_C}\\
            \end{bmatrix}$}, \\
            d_F&\in \mathbb{R}_{>0}, \,  S \in \C^{2\times2},\, S = S^H \succ 0 \bigg\} .\nonumber
    \end{align}
We can then use the approach of \cite{Janssen2024ModularPerspective} to formulate the following guarantee on the boundedness of $\ec(s)$.

\begin{thm} \label{thm:req_validation}
    Consider the transfer functions $\sys(s)$, $\env(s)$ and $\llft(\env,\sys)$ satisfying \autoref{ass1} and error dynamics \autoref{eq:ec_lft}. Let $W(s)$ and $V(s)$ be bistable and biproper weighting functions as given in \autoref{eq:DVW_tfs} and let the sets $\uees{22},\ \ueft$ and $\uec$ be as given in \autoref{eq:acc_specs}. If there exist some scaling matrices $(D_\ell,D_r)\in \mathbb{D}$, with $\mathbb{D}$ as given in \autoref{eq:D_def}, such that
    \begin{equation}\label{eq:req_LMI}\hspace{-1mm}
        \mathcal{N}(i\omega) D_r\mathcal{N}^H(i\omega)\preceq D_\ell \ \,\forall\ \omega\, \in\, \R, 
    \end{equation}
   with $\mathcal{N} = VNW$ and $N$ as in \autoref{eq:N_def}, then it follows that if 
   \begin{equation}\label{eq:ees_eft_req}
       \ees{22}\in \uees{22} \ \text{  and  }\ \eft\in \ueft, 
   \end{equation}
   $\llft(E,\sysr)$ is well-posed and internally stable and the coupled error dynamics satisfies
   \begin{equation}\label{eq:ec_req}
        \ec\in \uec.
    \end{equation}
\end{thm}
\begin{proof}
    We first note that $N\in\RH$ due to \autoref{ass1} and $\mathcal{N}\in\RH$ due to the weighting functions $V,\,W\in \RH$. Then, according to \cite[Theorem~3.2]{Janssen2024ModularPerspective}, for any $\ees{22},\, \eft \in \RH$ satisfying \autoref{eq:ees_eft_req}, the coupled error dynamics \autoref{eq:ec_lft} are well-posed, internally stable and satisfies \autoref{eq:ec_req} if and only if 
    \begin{equation}\label{eq:mu_req}
        \sup_{\omega\in\R}\ \mu_{\sud}\big(\mathcal{N}(i\omega)\big)< 1,
    \end{equation}
    where $\mu_{\sud}$ denotes the structured singular value \cite[Definition~3.1]{Packard1993TheValue}. Instead of $\mu_{\sud}$, we use its upper bound \begin{equation}\label{eq:mu_upper_bnd}
        \bar{\sigma}(D_\ell^{-1/2} \mathcal{N}(i\omega) D_r^{1/2}) \geq \mu_{\sud}(\mathcal{N}(i\omega)),    
    \end{equation}
    which can be tightened by optimizing over $D_\ell$ and $D_r$ \cite{Packard1993TheValue}. 
    
    Following \cite[Theorem~3.5]{Janssen2024ModularPerspective}, the requirement $\bar{\sigma}(D_\ell^{-1/2}\mathcal{N}(i\omega)D_r^{1/2})<1$ can be equivalently stated as $\mathcal{N}(i\omega) D_r\mathcal{N}^H(i\omega)\preceq D_\ell$. Therefore, \autoref{eq:req_LMI} implies \autoref{eq:mu_req} and thus $\ec\in \RH$ satisfies \autoref{eq:ec_req}. Due to the definition of $\ec$ in \autoref{eq:cpld_error_def} and \autoref{ass1}, this also guarantees the internal stability and well-posedness of $\llft(E,\sysr)$.
\end{proof}

\begin{rem}
    The approach of $\mu$-analysis, as exploited in \autoref{thm:req_validation}, typically requires $D_\ell,\, D_r$ to only satisfy $D_r^{1/2} \Delta = \Delta D_\ell^{1/2}$. However, we impose additional constraints on the positive definiteness and Hermitian form of $D_\ell,\, D_r$ in \autoref{eq:D_def}. These restrictions simplify further computation, while they do not increase the conservatism of \autoref{eq:mu_upper_bnd} \cite{Packard1993TheValue}.
    % The definition of $\mathbb{D}$ in \autoref{eq:D_def} also imposes $D_\ell$ and $D_r$ to be Hermitean, positive definite, and imposes the bottom right elements to equal the identity matrix. These added constraints further restrict $D_\ell$ and $D_r$ to simplify further computation, without affecting the outcome \cite{Packard1993TheValue}.
\end{rem}

Besides the relation of weighted $\Hnrm{\infty}$-norm \red{set definitions}, as in \autoref{thm:req_validation}, similar methods from robust performance are also applicable to relate frequency-dependent error bounds. To this end, with slight abuse of notation, we formulate frequency-dependent error \red{set definitions} by restricting $\ees{22}(i\omega)$, $\eft(i\omega)$ and $\ec(i\omega)$ to $\uees{22}(\omega)$, $\ueft(\omega)$ and $\uec(\omega)$ at frequency point $\omega \in \R$, where 
\begin{align} \nonumber 
        \uees{22}(\omega)&\coloneqq\Big\{\ees{22}(i\omega)\ \big|\\ \nonumber
        &\big\|\big(W_E(i\omega)\big)^{-1} \ees{22}(i\omega) \big(V_E(i\omega)\big)^{-1}\big\| \leq 1 \Big\}, \\ 
        \ueft(\omega) & \coloneqq\Big\{\eft(i\omega)\ 
        \big|   \label{eq:FD_acc_specs} \\ \nonumber
        &\big\|\big(W_F(i\omega)\big)^{-1} \eft(i\omega) \big(V_F(i\omega)\big)^{-1}\big\| \leq 1 \Big\}, \\ \nonumber
        \uec(\omega) & \coloneqq\big\{\ec(i\omega)\ \big| \ \|V_C(i\omega) \ec(i\omega) W_C(i\omega)\| < 1 \big\}.
\end{align}

Using the \red{set definitions} of \autoref{eq:FD_acc_specs}, we formulate a guarantee similar to \autoref{thm:req_validation} on the boundedness of $\ec(i\omega)$ at a specific frequency point $\omega \in \R$.
\begin{cor}\label{cor:freq_dep}
    Consider the transfer functions $\sys(s)$ and $\env(s)$, such that $\llft(\env,\sys)$ is well-posed and stable. Let $W(s)$ and $V(s)$ be weighting functions as in \autoref{eq:DVW_tfs}. Let $\omega \in \R$ and let the sets $\uees{22}(\omega),\ \ueft(\omega)$ and $\uec(\omega)$ be given as in \autoref{eq:FD_acc_specs}. If there exist $(D_\ell,D_r)\in \mathbb{D}$, with $\mathbb{D}$ as given in \autoref{eq:D_def}, such that
    \begin{equation}\label{eq:FD_req_LMI}\hspace{-1mm}
        \mathcal{N}(i\omega) D_r\mathcal{N}^H(i\omega)\leq D_\ell, 
    \end{equation}
   with $\mathcal{N}(i\omega) = V(i\omega)N(i\omega)W(i\omega)$ and $N$ as in \autoref{eq:N_def}, then it follows that if 
   \begin{equation}
       \ees{22}(i\omega)\in \uees{22}(\omega) \ \text{  and  }\ \eft(i\omega)\in \ueft(\omega), 
   \end{equation}
   the coupled error dynamics $\ec$, evaluated at $\omega$, satisfies
   \begin{equation}
        \ec(i\omega)\in \uec(\omega).
    \end{equation}
\end{cor}
\begin{proof}
    The proof closely follows \cite[Theorem~3.4]{Janssen2024ModularPerspective}, considering $\mu_{\sud}$ per frequency point, and is largely equivalent to \autoref{thm:req_validation}.
\end{proof}

%%%%%%%%%%%%%%%%%%%%%%%%%%%%%%%%%
\section{A priori specifications for abstracted reduction} \label{sec:specifications}
In this section, we will leverage the relation between the introduced errors, as determined in \autoref{sec:bnd_relations}, to guide the abstraction of $\env$ and reduction of $\sys$. This allows us to guarantee that the resulting coupled error $\ec$ satisfies a user-defined, possibly frequency-dependent specification. Specifically, we will answer the question: \emph{How should $\envr(s)$ and $\sysr(s)$ be selected, such that $\llft(\env,\sysr)$ is stable and meets a prescribed accuracy specification?} 

To this end, we adopt the perspective in which $\uees{22}$, $\ueft$ and $\uec$ represent specifications on $\ees{22}$, $\eft$ and $\ec$, respectively. Explicitly, an error, such as $\ec$, satisfies its specification, $\uec$, if $\ec\in\uec$. \autoref{thm:req_validation} and \autoref{cor:freq_dep} then allow us to relate such specifications. In particular, we focus on the case where $\uec$ is given and we mean to find specifications $\uees{22}$ and $\ueft$ that guarantee $\ec \in \uec$.

First, in \autoref{ssec:Hinf_specs}, we formulate an optimization problem where $\uec$ is given and $\uees{22}$ and $\ueft$, as in \autoref{eq:acc_specs}, are determined as the least conservative specifications on the abstraction and reduction errors. This result is subsequently used to formulate the robust abstracted reduction framework, to efficiently reduce $\llft(\env,\sys)$ to $\llft(\env,\sysr)$, such that $\llft(\env,\sysr)$ is well-posed and internally stable and $\ec\in \uec$.

Additionally, we formulate a frequency-dependent variant of the above-mentioned optimization problem in \autoref{ssec:FD_specs}, using \autoref{cor:freq_dep}, which is more flexible, but provides no stability guarantees. We then conclude in \autoref{ssec:considerations} with some notes on the properties of the proposed methods.

%%%%%%%%%%%%%
\subsection{The robust abstracted reduction framework for stability and accuracy guarantees}\label{ssec:Hinf_specs}

To guarantee the stability and accuracy of $\llft(\env,\sysr)$, captured in $\uec$, we will use \autoref{thm:req_validation} to provide suitable specifications $\uees{22}$ and $\ueft$. This can be interpreted as a distribution of the coupled error budget $\uec$ over the abstraction error budget $\uees{22}$ and the structure-preserving reduction error budget $\ueft$. To maximize the reduction of $\env$ and $\sys$, we aim for lenient specifications (large error budgets) $\uees{22}$ and $\ueft$, as in \autoref{eq:FD_acc_specs}. To this end, we prescribe the weighting matrices $W(s)$, as in \autoref{eq:DVW_tfs}, and $\bV(s)$, defined as
\begin{equation}\label{eq:bV_def}
   \begin{aligned}
       \bV(s) &\coloneqq \diag(\bV_\env(s),\bV_F(s),\bV_\env(s),\bV_C(s)),\\
       &=\eb^{-1} V(s),
    \end{aligned}         
\end{equation} 
where $\eb= \diag(\ebe I_p,\ebf I_m,\ebe I_p,I_{m_c})$ and $V$ is given in \autoref{eq:DVW_tfs}. Particularly, $\ebe,\ebf\in\R_{>0}$ represent $\Hnrm{\infty}$-bounds on the weighted errors as
\begin{equation} \label{eq:eb_def}
    \begin{gathered}
        \|W_E^{-1} \ees{22} \bV_E^{-1}\big\|_\infty \leq \ebe, \\
        \|W_F^{-1} \eft \bV_F^{-1}\big\|_\infty \leq \ebf
    \end{gathered}
\end{equation}
and we aim to maximize $\ebe$ and $\ebf$. Then, for a given $W(s)$ and $\bV(s)$, $\uees{22}$ and $\ueft$ are defined by $W(s)$ and $V(s) = \eb \,\bV(s)$, as in \autoref{eq:FD_acc_specs}.

It is important to note that there exist infinitely many combinations of $\ebe$ and $\ebf$ that guarantee $\ec \in \uec$. To select which unique combination is the ``optimal'' one, we introduce an additional scalar parameter $\beta$ to weigh the relative importance of $\ueft$ over $\uees{22}$. This results in the following optimization problem.
\begin{thm}\label{th:opt_alg_TD_infty}
    Consider the transfer functions $\sys(s)$, $\env(s)$ and $\llft(\env,\sys)$ satisfying \autoref{ass1} and error dynamics \autoref{eq:ec_lft}. Let the requirements $\uees{22},\ \ueft$ and $\uec$ be given as in \autoref{eq:acc_specs} and let $\bV(s)$ and $W(s)$ be prescribed bistable and biproper weighting functions as in \autoref{eq:bV_def} and \autoref{eq:DVW_tfs}, respectively.
    Consider now the optimization problem where $\beta\in\R_{>0}$ is a given tuning variable:
    \begin{align}\label{eq:opt_alg_TD_infty}
        \text{\emph{given} }\ & \bV(s),\,W(s)\\
        \text{\emph{maximize} }\ & \ebe^2 + \beta\, \ebf^2 \nonumber\\
        \text{\emph{subject to} }\ & 
        \eb^2 \, \mathcal{N}(i\omega) D_r \mathcal{N}^H(i\omega)\preceq D_\ell \ \,\forall\ \omega\, \in\, \R,  \nonumber\\
        &\ (D_\ell,D_r)\in \mathbb{D}, \nonumber
    \end{align}
    where $\mathcal{N} = \bV N W$, $N$ is given in \autoref{eq:N_def} and $\mathbb{D}$ in \autoref{eq:D_def}.

    If $\ebe$, $\ebf$ is a feasible solution to \autoref{eq:opt_alg_TD_infty}, then for any $\ees{22}\in \RH$ and $\eft\in \RH$ that satisfy
    \begin{equation}\label{eq:opt_alg_infty_spec}
        \ees{22} \in \uees{22} \text{ and } \eft \in \ueft,
    \end{equation}
    the reduced, interconnected system $\llft(\env,\sysr)\in \RH$ and the error system satisfies $\ec \in \uec$.
\end{thm}
\begin{proof}
    The proof follows from \autoref{thm:req_validation}, where only $V(s)$ is replaced by $\eb\,\bV(s)$. Substitution of $V = \eb\,\bV$ into \autoref{eq:req_LMI} gives the constraint of \autoref{eq:opt_alg_TD_infty}. 
\end{proof}

\begin{rem}
    The matrix inequality of the optimization problem in \autoref{th:opt_alg_TD_infty} is not linear in its unknowns, but can be solved efficiently by iteratively solving for $\eb$ and $D_r$, as shown in \cite{Janssen2023ModularApproach}.
\end{rem}

The solution to the optimization problem of \autoref{th:opt_alg_TD_infty} determines the maximum weighted $\Hnrm{\infty}$-bounds $\ebe$ and $\ebf$, that define $\uees{22}$ and $\ueft$. However, to make sure these accuracy specifications can be satisfied by significantly reduced models $\envr$ and $\sysr$, the weighting functions $\bV$ and $W$ should reflect the expected trend of the error system's frequency response. For example, if the reduction method for abstracting the environment is known to particularly approximate low-frequency dynamics, $\bV_E$ and $W_E$ could be prescribed as low-pass filters to ensure a more uniform response of $W_E^{-1}\ees{22}V_E^{-1}$. This results in a less conservative $\Hnrm{\infty}$-bound and a lower-order abstraction of $\env$. %Conversely, if $\bV_E$ and $W_E$ would be (foolishly) selected as high-pass filters, very little error is allowed at high frequencies, such that barely any abstraction of $\env$ is possible.

To systematically reduce an interconnected system, we present the \emph{robust abstracted reduction} framework: an extension of \autoref{alg:absred} with the optimization problem of \autoref{th:opt_alg_TD_infty}. Using this framework in combination with appropriate (structure-preserving) reduction methods, $\llft(\env,\sys)$ is efficiently reduced to $\llft(\env,\sysr)$, such that $\llft(\env,\sysr)$ is well-posed, internally stable and $\ec\in\uec$ (i.e., the reduced-order interconnected system satisfies a given error specification). 

\begin{alg} \label{alg:robabsred}
    \emph{Robust Abstracted Reduction}\\
    \textbf{Input:} Transfer functions $\sys(s)$, $\env(s)$ and $\llft(\env,\sys)$ satisfying \autoref{ass1}, bistable and biproper $\bV$ and $W$ as in \autoref{eq:bV_def} and \autoref{eq:DVW_tfs}, scalar $\beta \in \R_{\geq0}$ and full-rank weighting matrices $G_u \in \C^{m\times m}$ and $G_y \in \C^{p\times p}$.\\
    \textbf{Output:} Surrogate model $\sysr(s)$ of reduced order $r_\sys$, such that $\llft(\env,\sysr)$ is well-posed, internally stable and $\ec \in \uec$.
\begin{enumerate}
    \item \textbf{Optimization} Solve the optimization problem given in \autoref{th:opt_alg_TD_infty} to attain specifications $\ueft$ and $\uees{22}$.
    \item \textbf{Abstraction of $\env$}. Reduce $\env$ to $\envr$ of the lowest order $r_E$, such that $\ees{22}\in \uees{22}$.
    \item \textbf{Augmentation of $\envr$}. Augment the inputs and outputs of $\llft(\envr,\sys)$, resulting in $\llft(\fnvr,\sys)$, with $\fnvr$ as in \autoref{eq:F_defs}.
    \item \textbf{Reduction of $\sys(s)$}. Reduce $\sys(s)$ to $\sysr(s)$ of the lowest order $r_\sys$, such that $\eft \in \ueft$, where 
    \begin{equation}
    \eft = \llft\left(\left[\begin{smallmatrix}O &I\\ I & \envr_{22} \end{smallmatrix}\right], \sysr\right) - \llft\left(\left[\begin{smallmatrix}O &I\\ I & \envr_{22} \end{smallmatrix}\right], \sys\right).
    \end{equation}
    \item \textbf{Substitution of $\env$}. Substitute the original environment $\env$ to obtain $\llft(\env,\sysr)$.
\end{enumerate}
\end{alg}

%%%%%%%%%%%%%%%%%%%%%%%%%%%
\subsection{Frequency-based robust abstracted reduction }\label{ssec:FD_specs}

\autoref{alg:robabsred} fully addresses the problem statement of \autoref{ssec:prob_stat} by providing a systematic and efficient way to reduce $\llft(\env,\sys)$. However, the algorithm requires several inputs, among which the bistable weighting functions $\bV(s)$ and $W(s)$. In \autoref{ssec:Hinf_specs}, we advise $\bV(s)$ and $W(s)$ to reflect the expected trend error system's response, but this might be unknown or very irregular over the frequency range of interest. To avoid providing these weighting functions, we will formulate an alternative approach, based on \autoref{cor:freq_dep}, to relate the error specifications $\uees{22}(\omega),\ \ueft(\omega)$ and $\uec(\omega)$, as in \autoref{eq:FD_acc_specs}, for a specific discrete frequency grid $\omega \in \bbOmega$.

To simplify computation, we restrict $V(s)$ and $W(s)$, evaluated at $\omega$, to satisfy $V(i\omega) \in \mathbb{V}$ and $W(i\omega) \in \mathbb{W}$, with $\mathbb{V}$ and $\mathbb{W}$ given as    
    \begin{gather}\small
        \begin{aligned} \label{eq:cV_def}
           \hspace{-1mm} \mathbb{V} \coloneqq\Big\{& \diag( \cV_E,\cV_F,\cV_E,\cV_C) \Big| \\&\cV_E=\diag(v_{E,1},\shdots,v_{E,p})\in \R_{>0}^{p\times p}, \\ 
            &\cV_F=\diag(v_{F,1},\shdots,v_{F,m})\in \R_{>0}^{m\times m}, \\
            &\cV_C=\diag(v_{C,1},\shdots,v_{C,m_C})\in \R_{>0}^{m_C\times m_C}\Big\},
        \end{aligned}\\\small
        \begin{aligned} \label{eq:cW_def}
           \hspace{-1mm} \mathbb{W} \coloneqq\Big\{&\diag(\cW_E,\cW_F,\cW_E,\cW_C) \Big| \\&\cW_E=\diag(w_{E,1},\shdots,w_{E,m})\in \R_{>0}^{m\times m}, \\ 
            &\cW_F=\diag(w_{F,1},\shdots,w_{F,p})\in \R_{>0}^{p\times p}, \\
            &\cW_C=\diag(w_{C,1},\shdots,w_{C,p_C})\in \R_{>0}^{p_C\times p_C}\Big\}.
        \end{aligned}
    \end{gather}
This results in the following optimization problem.
\begin{thm}\label{th:opt_alg_TDw}
    Consider the transfer functions $\sys(s)$ and $\env(s)$, such that $\llft(\env,\sys)$ is well-posed and stable. Let $N(s)$, $\mathbb{D}$, $\mathbb{V}$ and $\mathbb{W}$, as given in \autoref{eq:N_def}, \autoref{eq:D_def}, \autoref{eq:cV_def} and \autoref{eq:cW_def}, respectively, and let the sets $\uees{22}(\omega),\ \ueft(\omega)$ and $\uec(\omega)$ be given as in \autoref{eq:FD_acc_specs}. Consider now the optimization problem at the frequency point $\omega\in\R$ where $V_C(i\omega)$ and $W_C(i\omega)$ are prescribed weighting matrices and $\beta\in\R_{>0}$ is a given tuning variable:
    \begin{align}\label{eq:opt_alg_TDw}
        \text{\emph{given} }\ & V_C(i\omega),\,W_C(i\omega)\\
        \text{\emph{minimize} }\ & \trace(X_V\, V^{-2}(i\omega)) + \trace(X_W\, W^{-2}(i\omega))\nonumber\\
        \text{\emph{subject to} }\ & \begin{bmatrix}\nonumber
            W^{-2}(i\omega)D_r^{-1} & N^H(i\omega) \\ N(i\omega) & V^{-2}(i\omega)D_\ell
        \end{bmatrix}\succ 0,\\\nonumber
        &V(i\omega) \in \mathbb{V}, \ W(i\omega) \in \mathbb{W}, \ (D_\ell,D_r)\in \mathbb{D},\\\nonumber\small
        &X_W = \diag(I_m,\beta I_p,I_m,I_{p_C}), \\ &X_V = \diag(I_p,\beta I_m,I_p,I_{m_C}). \nonumber
    \end{align}

    If $V(i\omega)$, $W(i\omega)$ is a feasible solution to \autoref{eq:opt_alg_TDw}, then for any $\ees{22}(i\omega)\in\uees{22}(\omega)$ and $\eft(i\omega)\in\ueft(\omega)$, it is ensured that $\ec(i\omega) \in \uec(\omega)$. 
\end{thm}
\begin{proof}
    In \cite[Theorem~1]{Janssen2023ModularApproach} the inequality of \autoref{eq:opt_alg_TDw} is shown to be equivalent to \autoref{eq:FD_req_LMI} when $V(i\omega) \in \mathbb{V}$ and $W(i\omega) \in \mathbb{W}$. The rest of the proof follows directly from \autoref{cor:freq_dep}.
\end{proof}
\begin{rem}
    The matrix inequality of the optimization problem in \autoref{th:opt_alg_TDw} is not linear in its unknowns, but can be solved efficiently by iteratively solving for $V^{-2}(\omega),\, W^{-2}(\omega)$ and $D_\ell,\,D_r$ as shown in \cite{Janssen2023ModularApproach}. 
\end{rem}

Compared to \autoref{th:opt_alg_TD_infty}, which required the bistable transfer matrices $\bV(s)$ and $W(s)$, \autoref{th:opt_alg_TDw} only needs to be supplied with the required coupled accuracy specification $\uec(\omega)$ for a specific frequency $\omega$. In addition, $\uec(\omega)$ can be selected more freely, as the weighting matrices $V(i\omega)$ and $W(i\omega)$ do not even need to represent actual transfer functions. However, this comes at the loss of any guarantees on well-posedness, stability or on an error bound for other frequency points.
\red{\begin{rem} A third abstracted reduction algorithm could be formulated based on \autoref{th:opt_alg_TDw}, using frequency-dependent bounds as in \autoref{eq:FD_acc_specs}. However, as this algorithm would be nearly equivalent to \autoref{alg:robabsred}, it is omitted for brevity. 
\end{rem}}
\vspace{-3mm}

%%%%%%%%%%%%%%%%%%%%%%%%%%%%%%%
\subsection{Computational considerations and tuning parameters}\label{ssec:considerations}

The main drawback of the systematic approach of \autoref{alg:robabsred} is the increase of computational cost with respect to standard abstracted reduction as in \autoref{alg:absred}. The additional cost, however, heavily depends on the system under consideration. The computational cost of the optimization problems as presented in \autoref{th:opt_alg_TD_infty} and \autoref{th:opt_alg_TDw}, scales with the number of inputs and outputs of the nominal model $N(s)$ \cite{Janssen2024ModularPerspective}. Therefore, the robust approach is best applicable to systems where the number of inputs and outputs, $m$ and $p$, are small (e.g., less than $10$). Furthermore, in the frequency-based approach, optimization is typically conducted for a discrete frequency grid $\omega \in \bbOmega$. To expedite this process, parallel computation can be employed, simultaneously addressing multiple frequency points.

% To reduce $\env(s)$ and $\sys(s)$ such that $\ees{22}\in\uees{22}$ and $\eft\in\ueft$, we employ the following subroutine.
% \begin{alg}\label{alg:red2bnd}
%     \emph{Reduction to meet error specification}\\
%     \textbf{Input:} System $\sys(s)$, and error specification $\ue = \{\Lambda\, \big|\,\|W\Lambda V\|\leq 1\}$.\\
%     \textbf{Output:} Surrogate model $\sysr(s)$ of reduced order $r_\sys$, such that $\sys(s)-\sysr(s) \in \ue(\omega)\ \forall \,\omega \in \bbOmega$.
%     \begin{enumerate}
%         \item Calculate the frequency response $H(\omega)=\sys(i\omega),\ \forall\, \omega\in\bbOmega$ and set $r_\sys = 0$.
%         \item Reduce $\sys(s)$ to $\sysr(s)$ of order $r_\sys$.
%         \item Calculate frequency response $\hat{H}(\omega)=\sysr(i\omega),\ \forall\, \omega\in\bbOmega$.
%         \item Set $r_\sys = r_\sys+1$ and repeat steps 2-4 until $\|W(\omega)(\hat{H}(i\omega)-H(\omega))V(i\omega)\|\leq 1,\ \forall \,\omega \in \bbOmega$.
%     \end{enumerate}    
% \end{alg}
Besides the additional computational cost of optimization in step 1) of \autoref{alg:robabsred}, the robust approach also complicates the reduction of $\env$ and $\llft(\fnvr,\sys)$ in step 2) and 4). Namely, the models should be reduced as much as possible, while satisfying $\ees{22}\in\uees{22}$ and $\eft\in\ueft$, which typically results in an iterative procedure. In case of standard, projection-based reduction methods, this too can be implemented efficiently by iteratively increasing the order of the projection matrix. To check efficiently whether $\ees{22}\in\uees{22}$ and $\eft\in\ueft$ at each iteration, the $\Hnrm{\infty}$-norm can be calculated using the approach of \cite{Bruinsma1990AMatrix}. If, instead, frequency-based bounds are used, as in \autoref{th:opt_alg_TDw}, $\env_{22}(i\omega)$ and $\llft\big(\fnvr(i\omega),\sys(i\omega)\big)$ can be determined beforehand, such that $\ees{22}(i\omega)$ and $\eft(i\omega)$ can be determined rapidly by only calculating $\envr_{22}(i\omega)$ and $\llft(\fnvr(i\omega),\sysr(i\omega)$. Particular reduction methods also permit a priori lower or upper error bounds \cite{Glover1984AllBounds,Liu1989SingularSystems}, which may be used to select an initial $r_E,\ r_\sys$ for faster convergence.

In \autoref{alg:robabsred}, $\beta$, $G_y$ and $G_u$ are tuning parameters to be supplied by the user. Although any selection of these parameters guarantees $\ec$ to be in $\uec$, they greatly influence the resulting reduced-order model $\sysr(s)$ and the required computational cost. Their specific influence is discussed below:

\noindent\textbf{Optimization parameter $\beta$}

Parameter $\beta$ is used to choose how to allocate the total allowed reduction error $\uec$ between $\uees{22}$ and $\ueft$ in the optimization of \autoref{th:opt_alg_TDw}. A larger $\beta$ therefore increases the set size of $\ueft$ at the cost of a smaller set size of $\uees{22}$, such that $\sys(s)$ can be reduced further and $\env(s)$ can be reduced less. Vice versa, a smaller $\beta$ allows more reduction of $\env(s)$ and lesser reduction of $\sys(s)$.
Thus, $\beta$ gives a trade-off between the computational cost of the structure-preserving reduction (smaller $\envr(s)$ reduces cost) and the amount of effective reduction of $\sys(s)$.
\begin{rem}\label{rem:beta and udft bnd}
    While $\beta$ allows the user to increase the size of $\ueft$, there exists an upper bound to $\ueft$ for which $\uees{22}$ is an empty set, i.e., $\ees{22} = O$. At this point, the reduction converts to the expensive structure-preserving reduction of $\llft(\env,\sys)$, which we aim to avoid with abstracted reduction (see also \autoref{ssec:error_charac}). For a sufficiently small $\uees{22}$ however, $\ueft$ approximates this upper bound and further increasing $\beta$ does not allow further reduction of $\sysr(s)$. This value of $\beta$ is in some sense optimal, as $\sysr(s)$ is reduced as far as possible using \autoref{alg:robabsred}, and further increasing $\beta$ only results in an increased order of $\envr(s)$ and thus in an increased computational cost of the reduction. This value is however only attainable through iterative evaluation of the resulting $r_\sys$ and $r_\env$ for variable $\beta$, which is very costly itself.
\end{rem}

\noindent\textbf{Augmented weighting $G_u$ and $G_y$}

Weighting matrices $G_u$ and $G_y$ scale the magnitude of the augmented outputs. Intuitively, increasing their magnitude therefore increases the importance of the augmented outputs, i.e., the 22-partition of $\llft(\fnvr,\sys)$ as in \autoref{eq:ef22_Def}, with respect to the external inputs and outputs of $\llft(\envr,\sys)$. Large $G_u$ and $G_y$ would therefore improve the approximation  $\llft(\fnvr,\sys)_{22}$, such that fewer states are required to achieve $\eft\in\ueft$, whereas small $G_u$ and $G_y$ would improve the approximation $\llft(\fnvr,\sys)_{11} = \llft(\env,\sys)$. Thus, the magnitudes of $G_u$ and $G_y$ give a trade-off between accuracy of $\llft(\env,\sysr)$ and the amount of reduction. This means that for smaller $G_u$ and $G_y$, $\ec$ becomes smaller, i.e., the reduction becomes more conservative with respect to the specified $\uec$, because 1) there is more focus on the accuracy of $\llft(\envr,\sys)$ compared to $\llft(\fnvr,\sys)_{22}$ and 2) $r_\sys$ is increased.

In addition to the simple scaling of magnitudes, $G_u$ and $G_y$ can also be related to $V_F(s)$ and $W_F(s)$, for instance, as
\begin{equation} \label{eq:WV_sys_inv}
    G_{u,ij} = \alpha \left(\|V_{F,ij}\|_2\right)^{-2}, \quad G_{y,ij} = \alpha \left(\|W_{F,ij}\|_2\right)^{-2},
\end{equation}
where $(.)_{ij}$ denotes the $i$,$j$'th element of the transfer matrix, $\alpha$ is a scalar $\R_{>0}$ and $\|(.)\|_2$ denotes the $\Hnrm{2}$-norm. By the inverse relation of \autoref{eq:WV_sys_inv}, the reduction can be steered to focus most on the input-output pairs for which the bounds are tight, allowing further reduction. Additional frequency dependencies can also be reflected by augmentation with transfer matrices $G_u(s)$ and $G_y(s)$, which should be bistable and biproper to allow \autoref{lem:error_corr}. This is considered out of the scope of this paper.
% \begin{rem}\label{rem:WV_sys and CLBR}
%     While $G_y$ and $G_u$ can theoretically guide the distribution of accuracy by influencing the structure-preserving reduction method, we have observed that ISBT does not respond intuitively to applied weighting with $G_y$ and $G_u$. In a general sense, ISBT does not consistently prioritize the approximation of inputs and outputs with highest magnitude, as expected from its rationale in \cite[Lemma~2]{Vandendorpe2008ModelSystems}. %This might be explained by the already extensive, dynamic weighting of $\sys(s)$ by $\envr(s)$, such that additional weighting using $G_y$ and $G_u$ is not straightforwardly observed. 
%     Further research might help to increase our understanding in this regard.
% \end{rem}

% The robust approach, as presented in \autoref{alg:robabsred}, is an effective method to translate a global accuracy specifications $\uec(\omega)$ into an appropriate reduced order model $\sysr(s)$. In addition, by reducing $\sys(s)$ using a structure-preserving reduction method, coupled to $\envr(s)$, $\sys(s)$ can be reduced further than with standard subsystem reduction. 

%
%%%%%%%%%%%%%%%%%%%%%%%%%%%%%%%%%%%%%%%%%
%
\section{Reduction of an interconnected, structural-dynamics model}\label{sec:case_study_red}
\begin{figure}
    \centering
    \includegraphics[width = \linewidth]{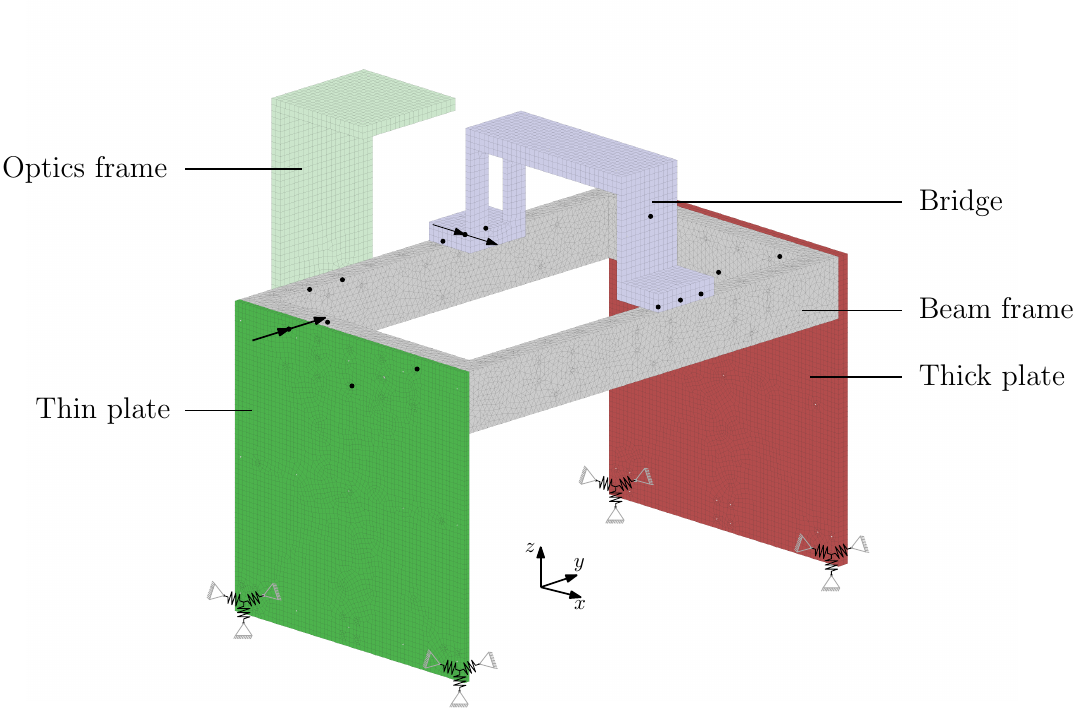}
    \caption{Schematic drawing of the benchmark system, where \groundconn \ indicates a spring connection to the ground, $\bullet$ indicates an interface point and \IOsymb\ indicates a collocated external input/output pair.}
    \label{fig:benchmark}
\vspace{-2mm}
\end{figure}

To evaluate the framework of abstracted reduction, we consider a structural-dynamics model for equipment in the lithography industry. The system, as shown schematically in \autoref{fig:benchmark}, represents a simplified, scaled-down frame of a lithography machine, consisting of several interconnected components. The system is a typical example of an interconnected structural dynamics system encountered in industry.

The interconnected system is about $25\!\times\! 42\!\times\! 40\,$cm ($x\times y\times z$) and is made of 5 aluminium components (or subsystems), as shown in \autoref{fig:benchmark}: two plates of different thickness, a frame for the support of optical equipment, a bridge-like frame and a frame of beams connecting all components. The two plates are connected to the ground with three translational springs with a stiffness of $10^{10}\,$N/m each and three rotational springs with a stiffness of $10^{5}\,$Nm/rad each in two locations per plate, see \autoref{fig:benchmark}. The components are connected to each other with linear springs with a stiffness of $10^8\,$N/m per spring in x-, y- and z-directions at the interface points shown in \autoref{fig:benchmark}. Each component has a modal damping factor \cite{Gawronski2004AdvancedStructures} of 5\% for each eigenmode. As a result, the interconnected model has no rigid-body modes and is asymptotically stable. 

\red{All components are} modeled by high-order transfer function matrices with forces as their inputs and collocated displacements as outputs, such that the spring interconnection is represented by a static interconnection matrix \cite{Janssen2023ModularApproach}. The model order and the number of inputs and outputs per component model are indicated in \autoref{tab:dimensions}. The interconnected system has two external, collocated, force inputs and displacement outputs ($m_C = p_C =2$): one at the interface between the thin plate and the beam frame in y-direction and one at the interface between the bridge and the beam frame in x-direction, as indicated in \autoref{fig:benchmark}.

\begin{table}
\caption{Model orders $n_\sys$, number of inputs and outputs $m=p$, order of the corresponding abstracted environment $r_E$, and order $r_\sys$ after reduction.}
\label{tab:dimensions}
\centering
\begin{tabular}{lcccc}\toprule
 Name             & $n_\sys$ & $m=p$ & $r_E$ & $r_\sys$ \\ \midrule
Thin plate       & 380   & 9     & 20 & 20\\
Thick plate      & 358   & 9     & 22 & 20\\
Beam frame       & 1038  & 45    & 90 & 32\\
Bridge           & 208   & 18    & 38 & 32\\
Optics frame     & 152   & 9     & 22 & 32\\ 
\hline
Interconnected system & 2136  & 2  & - & 100\\ \bottomrule
\end{tabular}
\vspace{-2mm}
\end{table}

\red{The original interconnected system model, with an order of 2136, is too computationally costly for various industrial applications, such as real-time simulation, (low-order) controller synthesis, online monitoring, and sensitivity analysis. Therefore, in this section, we will reduce the interconnected system model to a lower order. Specifically, we will perform two evaluations:}
\begin{enumerate}
    \item In \autoref{ssec:fixed_order_red}, we evaluate the abstracted reduction framework of \autoref{alg:absred}, where we select the reduction orders $r_E$ and $r_\sys$ beforehand, as indicated in \autoref{tab:dimensions}, and reduce all subsystems.
    \item In \autoref{ssec:robust_routine}, we evaluate the frequency-based robust abstracted reduction framework of \autoref{ssec:FD_specs}, where we input $\uec(\omega)$ and reduce the thin plate to an order $r_\sys$, such that the resulting $\ec(s) \in \uec(\omega),$ for all $\omega \in \bbOmega$.
\end{enumerate}

The abstracted environment $\envr(s)$ is obtained by the Hintz-Herting (HH) component mode synthesis method \cite{Herting1985AMethod}, whereas the reduction of $\sys(s)$ (a component model) is performed using Closed-Loop Balanced Reduction (CLBR), as introduced by Ceton and Wortelboer et al. \cite{Ceton1993FrequencyReduction,Wortelboer1994Frequency-weightedTools}. Specifically, the reduction of $\sys(s)$ is performed using residualization (also called singular perturbation), which retains the steady state transfer. See \cite{Antoulas2005ApproximationSystems} for more details.

\subsection{Fixed-order reduction} \label{ssec:fixed_order_red}
For this evaluation, all $5$ subsystems are reduced to a lower order $r_\sys$, as indicated in \autoref{tab:dimensions}, using the abstracted reduction approach introduced in this paper. To this end, we first formulate an environment model $\env_j$, for each $j\in\{1,\dots,5\}$, as the interconnection of all remaining subsystems $\sys_l$, $l\in \{1,\dots,5\}\backslash \{j\}$, and then perform \autoref{alg:absred} once for each subsystem $\sys_j$. 

To properly evaluate the reduction of all $\sys_j$ in the abstracted reduction framework, three balanced reduction methods are used: subsystem balanced reduction (ssBR), without taking the environment into account (see \red{\autoref{fig:ssRed_RGB} and} \cite{Moore1981PrincipalReduction}), interconnected systems balanced reduction (ISBR), taking its full environment into account (see \red{\autoref{fig:spRed_RGB} and} \cite{Vandendorpe2008ModelSystems}), and abstracted reduction using CLBR (aCLBR), taking only an abstraction of the environment into account (see \red{\autoref{fig:absRed_RGB}, }\autoref{alg:absred} and \cite{Wortelboer1994Frequency-weightedTools}). The three resulting reduction methods are then compared in terms of computational cost and the accuracy of the resulting reduced-order, interconnected model.

\begin{figure*}\centering
    \subfloat[\small\label{fig:FRF1}]{%
      \includegraphics[width=0.43 \linewidth]{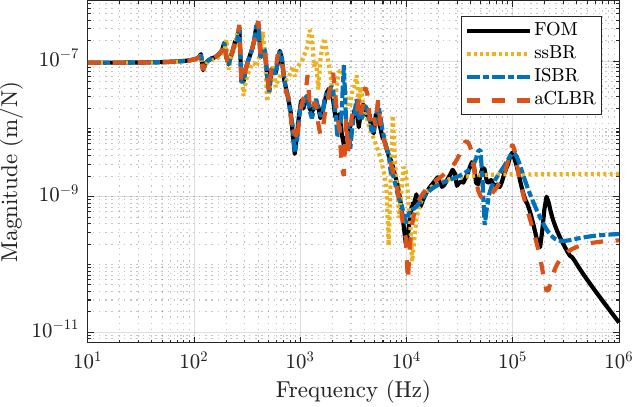}
    }\hspace{10mm}
    \subfloat[\small \label{fig:FRF2}]{%
      \includegraphics[width=0.43\linewidth]{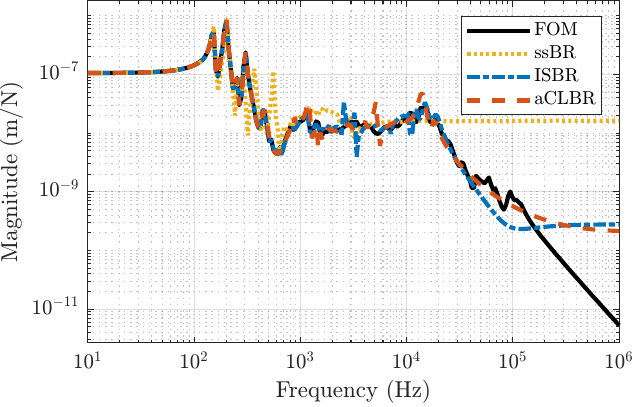}
    }\\
    \subfloat[\small\label{fig:eFRF1}]{%
      \includegraphics[width =0.43\linewidth]{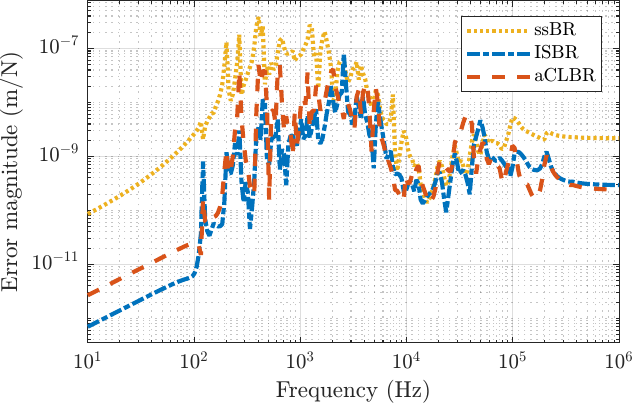}
    }\hspace{10mm}
    \subfloat[\small \label{fig:eFRF2}]{%
      \includegraphics[width=0.43\linewidth]{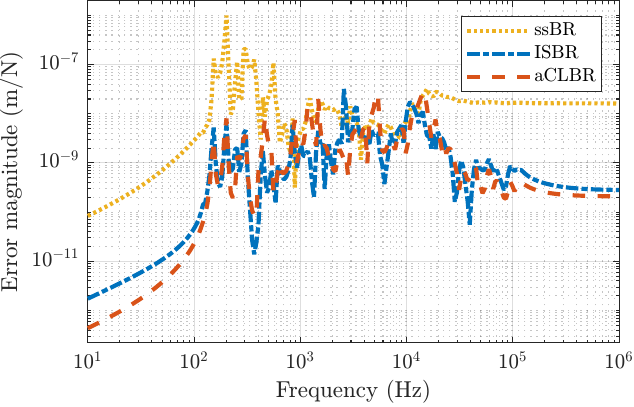}
    }
    \caption{The collocated FRFs of the benchmark model at the interface on the bridge (a) and the thin plate (b), and the FRFs of the corresponding error entries from $\ec$ in (c) and (d), respectively. The full-order model (FOM, \protect\FOMline) is compared to the models reduced by ssBR (\protect\ssBRline), ISBR (\protect\ISBRline) and aCLBR (\protect\aCLBRline)
    .}
    \label{fig:FRFs}
    \vspace{-3mm}
\end{figure*}

The ISBR approach, as introduced by Vandendorpe and Van Dooren \cite{Vandendorpe2008ModelSystems}, is a generalization of CLBR. Whereas CLBR is only defined for 2 interconnected subsystems, ISBR reduces all $k\geq2$ subsystems to retain their relevance with respect to the interconnected system. For aCLBR, we use $G_y = G_u = O_{m\times m}$, i.e., no augmentation, to focus solely on the accuracy of $\llft(\env,\sysr)$. 

The accuracy of the reduced-order, interconnected models is measured by visually comparing their frequency response functions (FRFs) with the FRFs of the unreduced interconnected model, the FRFs of their error $\ec(s)$ and the bounded $\Lnrm{2}$-norm of $\ec(s)$ \citep{Zhou1998EssentialsControl} (on the frequency-domain of $10$ to $10^6\,$Hz, relevant for this application) given by
\begin{equation}
    \|\ec\|_2^2 = \frac{1}{2\pi} \int_{2\pi\times 10}^{2\pi\times 10^6} \text{trace}[\ec^H(j\omega)\ec(j\omega)] \text{d}\omega.
\end{equation}

\autoref{tab:redmeds_characs} compares the computational cost of the reduction methods\footnote{All computation is performed on a HP Z-book Studio G5 with a Intel(R) Core(TM) i7-9750H CPU and 16 GB of RAM.} and the obtained accuracy. Clearly, and as expected, ssBR is the most computationally efficient, as only the dynamics of the individual subsystem models are evaluated upon their reduction. ISBR, on the other hand, considers the full interconnected model dynamics, which results in a considerable increase in computational cost with a factor 12. By taking only an abstraction into account, aCLBR reduces this cost by a factor 6 to only twice the cost of ssBR.

\begin{table}
\caption{Computation time and accuracy of the reduction methods.}
\label{tab:redmeds_characs}
\centering
\begin{tabular}{lccc} \toprule
Method                 & Time (s) & $\|\ec\|_2$ (m/N)   & \red{$\frac{\|\ec\|_2}{\|\llft(E,\Sigma)\|_2}$ (-)}\\ \midrule
ssBR                   & 19       & 2.56$\times10^{-5}$ & \red{1.45} \\
Full ISBR              & 231      & 6.24$\times10^{-6}$ & \red{0.356}\\
Abstracted CLBR        & 39       & 7.70$\times10^{-6}$ & \red{0.439} \\ \bottomrule 
\end{tabular}
\vspace{-4mm}
\end{table}

\begin{rem}
    When $\sys(s)$ represents the beam frame, the order of $\llft(\fnvr,\sys)$ is 1128, which remains relatively high. This high order can be attributed to the disparity in order between the beam frame and the other subsystems. Recall that aCLBR demonstrates its greatest efficiency improvement when the environment is substantially larger than the subsystem, as illustrated in \autoref{fig:cost reduction}. If all subsystems would be of similar size, the computational cost of aCLBR would approach the cost of ssBR.
\end{rem}

As mentioned before, \autoref{tab:redmeds_characs} also indicates the accuracy of the various reduced-order models (ROMs) by means of the \red{(relative)} $\Lnrm{2}$-norm of the corresponding error systems. \red{While all relative $\Lnrm{2}$-norms are quite large due to the significant order reduction, a clear distinction in accuracy is observable. First of all, the} ssBR-ROM results in a significantly higher $\Lnrm{2}$ error norm than the other two ROMs. This is as expected, as ssBR considers only the separate subsystems. Furthermore, the accuracy of the ISBR-ROM, in terms of its $\Lnrm{2}$ error norm, shows a slight superiority with respect to the accuracy of the aCLBR-ROM. This small sacrifice in accuracy is a logical result from the introduced abstraction error with aCLBR. We can conclude that aCLBR is capable of achieving an accuracy level comparable to ISBR, while requiring computational cost almost competitive wth ssBR.

The magnitudes of the two collocated FRFs of the full-order model (FOM) of the interconnected system are visualized together with the FRFs of the three corresponding reduced-order models (ROMs) in Figures \ref{fig:FRF1} and \ref{fig:FRF2}. The FRFs of the collocated errors from $\ec(s)$, i.e., the differences between the FRFs of ROMs and of the FOM, are visualized underneath in Figures \ref{fig:eFRF1} and \ref{fig:eFRF2}. These error plots reflect in more detail the observations made above regarding the $\Lnrm{2}$ error norms. Namely, the ssBR-ROM displays a clearly inferior accuracy with respect to the other two, whereas the accuracy of the ISBR- and aCLBR-ROMs seems similar.

Summarizing, we have evaluated \red{(open-loop) subsystem reduction (ssBR)}, structure-preserving reduction (ISBR) and an abstract implementation (aCLBR), as presented in \autoref{alg:absred}, by means of a benchmark system from the field of structural dynamics. For this example, the proposed abstracted reduction approach achieves similar accuracy to the accurate but expensive ISBR method, while having a computational cost similar to the cheap but inaccurate ssBR method.

\vspace{-2mm}
\subsection{Robust routine for abstracted reduction} \label{ssec:robust_routine}
In this example, we will focus on the reduction of the thin plate, denoted $\sys(s)$, such that the remaining (connected) components/subsystem models form the environment model $\env(s)$ with an order of $n_E = 1756$, $p+m_C = 11$ inputs and $m+p_C = 11$ outputs. Our goal is to reduce $\sys(s)$ to $\sysr(s)$ with abstracted reduction, such that $\ec(i\omega) \in\uec(\omega)$ for all $\omega \in \bbOmega$, where $\ec\coloneqq\llft(\env,\sysr)-\llft(\env,\sys)$. 

\begin{figure}\centering
    \subfloat[\small\label{fig:FRF_TDg}]{%
      \includegraphics[width=.8\linewidth]{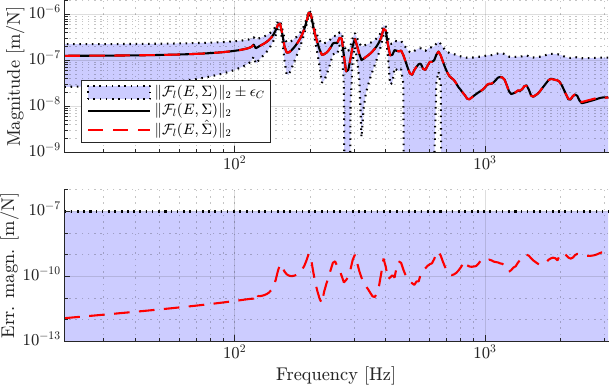}
    }\\
    \subfloat[\small \label{fig:FRF_TDe}]{%
      \includegraphics[width=.8\linewidth]{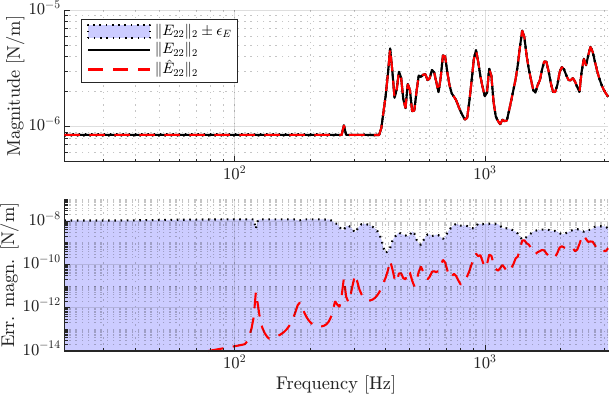}
    }\\
    \subfloat[\small\label{fig:FRF_TDf}]{%
      \includegraphics[width =.8\linewidth]{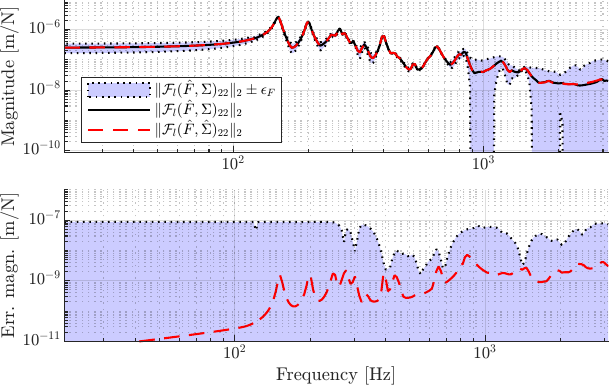}
    }
    \caption{Spectral norms of the MIMO transfer functions $\llft(\env,\sys)$ (a), $\env_{22}$ (b) and $\llft(\fnvr,\sys)_{22}$ (c), their reduced counterparts and corresponding bounds $\varepsilon_C(\omega)$, $\varepsilon_E(\omega)$ and $\varepsilon_F(\omega)$, respectively. The full-order models are given in black (\textcolor{black}{\textbf{---}}), whereas the reduced models are given in red (\textcolor{ML_c2}{\textbf{--\,--}}) and the error bounds in blue (\,\errboundbox\,).} 
    \label{fig:TD_FRFs}
    \vspace{-4mm}
\end{figure}

We first select the frequency grid $\bbOmega$ as 250 logarithmically equidistantly spaced points between 5\,Hz and 2\,kHz. We then visualize the FRF matrices of $\llft(\env,\sys)$, $\env_{22}(s)$ and $\llft(\fnvr,\sys)$, by means of their spectral norm per frequency in $\bbOmega,$ as the black lines in \autoref{fig:FRF_TDg}, \autoref{fig:FRF_TDe} and \autoref{fig:FRF_TDf}, respectively. 

To simplify interpretation of the accuracy specifications, we restrict the scaling matrices to be of the form
\begin{equation}
    \begin{aligned}
        W_E(i\omega) &= \sqrt{\varepsilon_E(\omega)} I_m,& V_E(i\omega) &= \sqrt{\varepsilon_E(\omega)} I_p, \\
        W_F(i\omega) &= \sqrt{\varepsilon_F(\omega)} I_p, & V_F(i\omega) &= \sqrt{\varepsilon_F(\omega)} I_m,\\
        W_C(i\omega) &= \sqrt{\varepsilon_C(\omega)} I_{p_C},& V_C(i\omega) &= \sqrt{\varepsilon_C(\omega)} I_{m_C},
    \end{aligned}    
\end{equation}for nonnegative, real scalars $\varepsilon_E(\omega),\, \varepsilon_F(\omega),\, \varepsilon_C(\omega)$. The error specifications can then be simplified to 
\begin{equation}
    \begin{aligned}
        \uees{22}(\omega) &= \{\ees{22}(i\omega)\,|\ \|\ees{22}(i\omega)\| < \varepsilon_E(\omega)\},\\
        \ueft(\omega) &= \{\eft(i\omega)\,|\ \|\eft(i\omega)\| < \varepsilon_F(\omega)\},\\
        \uec(\omega) &= \{\ec(i\omega)\,|\ \|\ec(i\omega)\| < \varepsilon_C(\omega)\}.
    \end{aligned}
\end{equation}

We specify the allowed coupled inaccuracy $\uec(\omega)$ by setting $\varepsilon_C(\omega) = 10^{-7},$ for all $\omega \in \bbOmega$. The coupled inaccuracy specification $\uec(\omega)$ is visualized accordingly in \autoref{fig:FRF_TDg} as the blue area. \red{This rather simple choice of $\varepsilon_C(\omega)$ emphasizes the accurate prediction of the resonance peaks, which might be relevant in, for example, a structural health monitoring use case where structural wear or damage may result in a frequency shift of the resonance peaks.} The specifications $\uees{22}(\omega)$ and $\ueft(\omega)$ are subsequently determined using \autoref{th:opt_alg_TDw} for a range of $\beta$. 

\begin{figure}[t]
    \subfloat[\small\label{fig:beta_bnds}]{%
      \includegraphics[width=0.47 \linewidth]{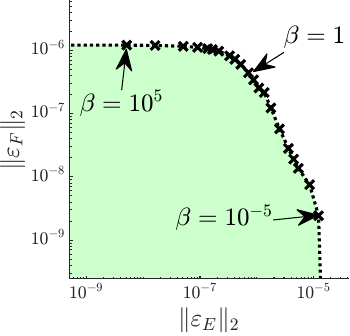}
    }
    \hfill
    \subfloat[\small \label{fig:beta_rs}]{%
      \includegraphics[width=0.47\linewidth]{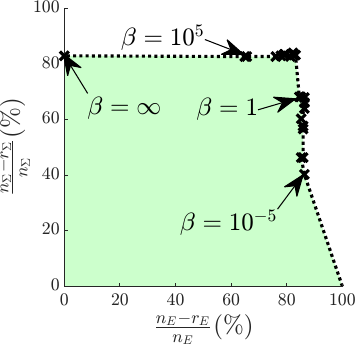}
    }
    \caption{(a) Euclidean norms of the error bounds $\varepsilon_E(\omega)$ and $\varepsilon_F(\omega)$ as determined from \autoref{th:opt_alg_TDw}, for different values of $\beta$, and (b) the maximum reduction allowed for the environment model $\env(s)$ and thin plate model $\sys$ to satisfy these bounds, respectively, using $G_y = I_m,\ G_u= I_p$.} \vspace{-1mm}
    \label{fig:beta}
\end{figure}

The influence of $\beta$ on the bounds $\varepsilon_F$ and $\varepsilon_E$ is visualized in \autoref{fig:beta_bnds}. As explained in \autoref{ssec:considerations}, $\beta$ influences the distribution of the total error budget $\uec(\omega)$. Specifically, a large $\beta$ increases the bound $\varepsilon_F(\omega)$ at the cost of a smaller bound $\varepsilon_E(\omega)$. For $\beta \rightarrow \infty$, only $\varepsilon_F(\omega)$ is maximized, regardless of $\varepsilon_E(\omega)$, i.e., $\varepsilon_F(\omega)$ is the bound on $\eft(\omega)$ if the full $\env(s)$ is considered instead of $\envr(s)$. For $\beta = 0$, the opposite occurs, and $\varepsilon_E(\omega)$ is maximized, regardless of $\varepsilon_F(\omega)$. However, the latter optimization problem is ill-posed as the $\uees{22}$ is theoretically unbounded for an empty set $\ueft$, because for $\eft = O$, $\sysr(s) = \sys(s)$ regardless of $\envr(s)$.

Naturally, a larger allowed error bound can be satisfied more easily by a reduced model, so that the model's order can be decreased further. Therefore, varying $\beta$ also allows a trade-off between $r_\sys$ and $r_E$. Assuming $G_y = I_p,\ G_u = I_m$, we can visualize this trade-off in \autoref{fig:beta_rs}. Note that we want to reduce $\sys(s)$ as far as possible, i.e, $n_\sys-r_\sys$ should be large, while $n_E-r_E$ should only be sufficiently small to reduce the computational complexity of the reduction approach. \autoref{fig:beta_rs} indicates that $\env(s)$ can be reduced by over 80\% without significantly influencing the allowed reduction of $\sys(s)$. This exemplifies the motivation of using environment abstraction for effective reduction of coupled systems.

\begin{figure}[t]
    \centering
    \includegraphics[width = 0.6\linewidth]{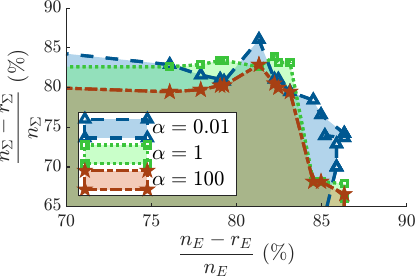}
    \caption{Maximum reduction allowed for the environment model $\env(s)$ and thin plate model $\sys$ to satisfy $\uees{22}(\omega)$ and $\ueft(\omega)$, respectively, for different weighting matrices $G_u =\alpha I_p$ and $G_y = \alpha I_m$.}
    \label{fig:alphabeta} \vspace{-6mm}
\end{figure}

In addition to the influence of $\beta$ on $\uees{22}$ and $\ueft$, weighting matrices $G_y$ and $G_u$ determine $\fnvr(s)$, enabling to change the importance of the augmented outputs compared to the interconnected system outputs in the structure-preserving reduction of $\llft(\fnvr,\sys)$ (see \autoref{ssec:considerations}). To evaluate the influence of these weighting matrices, we determine the maximum order reduction with weighting $G_y = \alpha I_p,\ G_u = \alpha I_m$ for $\alpha = 0.01,\ 1$, and $100$, as visualized in \autoref{fig:alphabeta}. As explained in \autoref{ssec:considerations}, a high value $\alpha$ favors the accuracy of the augmented outputs, such that, for a large $\alpha$, $\eft(s)$ would be relatively small, whereas $\ec(s)$ would be relatively large. This implies that for a given $r_\env$, a smaller $r_\sys$ would be required for $\eft\in \ueft$ if $\alpha$ is large. However, the results shown in \autoref{fig:alphabeta} are not consistent with this, as a larger $\alpha$ often allows less reduction of $\sys(s)$, given a certain given $r_\env$. In addition, we observe a rather nonlinear relation between error bounds and reduction order, which makes it difficult to set proper $\alpha$ and $\beta$ that result in the maximum reduction of $\env(s)$ and $\sys(s)$. This inconsistency between the theory and the results of \autoref{fig:alphabeta} is attributed to \red{the presented use case (where the dynamics of $\sys$ relevant to the input-output behaviour of $\sys$ and $\llft(E,\sys)$ seems to align) and the} non-optimality of CLBR (or weighted CLBR \cite{Wortelboer1994FrequencyConfiguration}). Further study is required to asses the influence of $G_y$ and $G_u$ \red{in alternate use cases} using other reduction methods.


Because the above evaluation of $\beta$ and $\alpha$ to find `optimal' values is usually too time-consuming \red{($\pm1$ hour for this use case)}, it is typically wise to select $\beta>1$, which puts emphasis on reduction of $\sys(s)$, accepting a higher order $\envr(s)$, resulting in a slightly higher computational cost (see \autoref{ssec:considerations}). For the remainder of the evaluation in this section, we will use $\beta = 100$. As CLBR does not respond intuitively to the weighting of $G_y$ and $G_u$, we simply select $G_y = I_p$, $G_u = I_m$. 

The frequency-dependent accuracy specifications $\uees{22}(\omega)$ and $\ueft(\omega)$ for these choices of $\beta,\ G_y$ and $G_u$ are visualized by the blue areas in \autoref{fig:FRF_TDe} and \autoref{fig:FRF_TDf}, respectively. The environment model $\env(s)$ and augmented, abstracted system $\llft(\fnvr,\sys)$ are subsequently reduced as far as possible subject to $\ees{22}(i\omega) \in \uees{22}(\omega), \ \eft(i\omega)\in \ueft(\omega),$ for all $\omega \in \bbOmega$ using Hintz-Herting reduction for the environment and CLBR for the thin plate model, resulting in $r_E = 184$ and $r_\sys = 63$ ($n_E = 1756$, $n_\sys = 380$). The FRFs of $\envr_{22}$ and $\llft(\fnvr,\sysr)_{22}$ and their errors are visualized by red dashed lines in \autoref{fig:FRF_TDe} and \autoref{fig:FRF_TDf}, respectively, which are, by definition, fully included in the blue area. Interconnection of $\sysr(s)$ with $\env(s)$ gives the reduced interconnected model $\llft(\env,\sysr)$, whose FRF and error are visualized by red dashed lines in \autoref{fig:FRF_TDg}, which are also well within the specification of $\uec(\omega)$.

\begin{rem}
    The full execution of the robust abstracted reduction framework of \autoref{alg:robabsred} using these settings takes approximately $450\,s$, of which most time is required to solve the optimization problem of \autoref{th:opt_alg_TDw}. \red{The computation time depends on the number of inputs and outputs of $\sys$ (which determine the dimensions of $N$) and the efficiency of the used numerical method.} Currently, $V^{-2}(i\omega),\ W^{-2}(i\omega)$ and $D_\ell,\ D_r$ are solved iteratively using MOSEK \cite{mosek}. However, there exist much more efficient methods to calculate bounds on $\mu_\Delta(VNW)$ (as used in \autoref{th:opt_alg_TD_infty}) \cite{Young1993RobustnessUncertainty}. Incorporating such methods could potentially speed up the determination of $\uees{22}$ and $\ueft$ significantly. This is, however, outside the scope of this paper and is left for further research.
\end{rem}

The robust abstracted reduction framework allocates an error budget, $\varepsilon_E(\omega)$, to the reduction of $\env$ and, $\varepsilon_F(\omega)$, to the reduction of $\sys$. To reduce $\env$ and $\sys$ as far as possible, we should aim to use as much of these budgets as possible. However, the chosen reduction methods do not take these error budgets into account, resulting in considerable conservatism. This is especially apparent for the abstraction, where $\|\ees{22}(i\omega)\| \,\red{\ll}\, \varepsilon_E(\omega)$ for most frequencies, as shown in \autoref{fig:FRF_TDe}. In addition, even when  $\|\ees{22}(i\omega)\| \approx \varepsilon_E(\omega)$, usually only one input-output pair restricts further reduction, while other input-output pairs are well within specification. 

These two sources of conservatism result in significant conservatism of the eventual reduced, interconnected model $\llft(\env,\sysr)$, as shown by difference of a factor 100 between actual error and allowed error. In addition, the reduced orders of $r_E=184$ and $r_\sys=63$ remain quite large. Through iterative reduction and evaluation of $\ec(s)$, a process that is far from straightforward, we found that the specification of $\uec(\omega)$, with $\varepsilon_C(\omega) = 10^{-7},$ for all $\omega \in \bbOmega$, is achievable by aCLBR with $r_E = 20$ and $r_\sys = 8$ and by ssBR with $r_\sys=22$. 

Summarizing, the above analysis gives insight into the sensitivity of the resulting error $\ec(s)$ to inaccuracies of the environment model. Namely, the environment model can be reduced significantly without seriously impacting the resulting accuracy of the reduced interconnected system model. This observation therefore also strengthens the case for using the general abstracted reduction framework of \autoref{alg:absred}.

\vspace{-1.5mm}
\subsection{Key insights from numerical evaluation}
For the considered use case, abstracted model reduction, as introduced in \autoref{sec:absred_framework} and evaluated in \autoref{ssec:fixed_order_red}, yielded an accuracy of the reduced interconnected system similar to the expensive structure-preserving method, while having a computational cost similar to the cheap subsystem reduction method. In other words, the framework of abstracted model reduction can significantly improve the computational tractability of structure preserving methods while retaining similar accuracy. This indicates that the low-order abstractions of the environment models are sufficient to indicate what subsystem dynamics needs to be retained to attain an accurate reduced-order interconnected system model.

This observation regarding the sufficiency of low-order abstractions is reaffirmed in the assessment of the robust abstracted reduction framework, as introduced in \autoref{sec:specifications} and evaluated in \autoref{ssec:robust_routine}, which showed that reducing the environment up to 80\% does not significantly impact the reduced system. Although the evaluation in \autoref{ssec:robust_routine} indicates that robust abstracted reduction tends to be conservative, this tendency might significantly mitigated by utilization of alternative reduction methods. In addition, the stability guarantee, given by the robust abstracted reduction framework, remains valuable even in cases of conservatism.

%
%%%%%%%%%%%%%%%%%%%%%%%%%%%%%%%%%%%%%%%%%
%
\section{Conclusion} \label{sec:con} 
We have introduced the framework of abstracted reduction to improve the tractability of the structure-preserving reduction of interconnected systems. In this framework, the structure-preserving reduction method is not applied to the full interconnected system model, but to a single subsystem model connected to a low-order abstraction of its environment. By thus applying structure-preserving reduction to a much lower-order model, the computational cost of this reduction is significantly reduced. Leveraging techniques from robust performance, we introduced a second approach to also automatically determine (environment) abstraction and (subsystem) reduction orders that guarantee the reduced interconnected system model to be stable and to satisfy a prescribed $\Hnrm{\infty}$-accuracy specification.

By means of a high-dimensional model representing an interconnected lithography machine frame, we have illustrated that employing a structure-preserving reduction method within the abstracted reduction framework can significantly decrease the reduction's computational cost without significant loss of accuracy. Although the robust performance-based approach tends to yield a conservative reduced-order model of the interconnected system, it effectively illustrates the relation between the introduced error budgets. It is important to emphasize that the introduced abstracted reduction framework is versatile, i.e., it is compatible with any structure-preserving reduction method. This versatility suggests the potential for numerous alternative implementations that offer enhanced accuracy, greater order reduction, and less conservatism compared to the current implementation.

%
%%%%%%%%%%%%%%%%%%%%%%%%%%%%%%%%%%%%%%%%%
%\
\section{Acknowledgements} \label{sec:ack}
The authors would like to thank Dr. Victor Dolk and Thijs Verhees, M.Sc., for valuable discussions.
\small{
\bibliographystyle{IEEEtran}
\bibliography{references} }

\begin{thebibliography}{10}
\providecommand{\url}[1]{#1}
\csname url@samestyle\endcsname
\providecommand{\newblock}{\relax}
\providecommand{\bibinfo}[2]{#2}
\providecommand{\BIBentrySTDinterwordspacing}{\spaceskip=0pt\relax}
\providecommand{\BIBentryALTinterwordstretchfactor}{4}
\providecommand{\BIBentryALTinterwordspacing}{\spaceskip=\fontdimen2\font plus
\BIBentryALTinterwordstretchfactor\fontdimen3\font minus
  \fontdimen4\font\relax}
\providecommand{\BIBforeignlanguage}[2]{{%
\expandafter\ifx\csname l@#1\endcsname\relax
\typeout{** WARNING: IEEEtran.bst: No hyphenation pattern has been}%
\typeout{** loaded for the language `#1'. Using the pattern for}%
\typeout{** the default language instead.}%
\else
\language=\csname l@#1\endcsname
\fi
#2}}
\providecommand{\BIBdecl}{\relax}
\BIBdecl

\bibitem{Dorosti2014FiniteControl}
M.~Dorosti, R.~H.~B. Fey, M.~F. Heertjes, M.~M.~J. van~de Wal, and
  H.~Nijmeijer, ``{Finite Element Model Reduction and Model Updating of
  structures for Control},'' \emph{IFAC Proceedings Volumes}, vol.~47, no.~3,
  pp. 4517--4522, 2014.

\bibitem{Baldwin2006ModularitySystems}
C.~Y. Baldwin and K.~B. Clark, ``{Modularity in the Design of Complex
  Engineering Systems},'' in \emph{Complex Engineered Systems. Understanding
  Complex Systems}, D.~Braha, A.~Minai, and Y.~Bar-Yam, Eds.\hskip 1em plus
  0.5em minus 0.4em\relax Springer, Berlin, Heidelberg, 2006, ch.~9, pp.
  175--205.

\bibitem{Reis2008ASystems}
T.~Reis and T.~Stykel, ``{A Survey on Model Reduction of Coupled Systems},'' in
  \emph{Model Order Reduction: Theory, Research Aspects and Applications},
  W.~H.~A. Schilders, H.~A. van~der Vorst, and J.~Rommes, Eds.\hskip 1em plus
  0.5em minus 0.4em\relax Berlin, Heidelberg: Springer Berlin Heidelberg, 2008,
  pp. 133--155.

\bibitem{Besselink2013AControl}
B.~Besselink, U.~Tabak, A.~Lutowska, N.~Van De~Wouw, H.~Nijmeijer, D.~J. Rixen,
  M.~E. Hochstenbach, and W.~H. Schilders, ``{A comparison of model reduction
  techniques from structural dynamics, numerical mathematics and systems and
  control},'' \emph{Journal of Sound and Vibration}, vol. 332, no.~19, pp.
  4403--4422, 2013.

\bibitem{DeKlerk2008GeneralTechniques}
D.~De~Klerk, D.~J. Rixen, and S.~N. Voormeeren, ``{General framework for
  dynamic substructuring: History, review, and classification of techniques},''
  \emph{AIAA Journal}, vol.~46, no.~5, pp. 1169--1181, 2008.

\bibitem{Antoulas2005ApproximationSystems}
A.~C. Antoulas, \emph{{Approximation of Large-Scale Dynamical Systems}}, R.~C.
  Smith, Ed.\hskip 1em plus 0.5em minus 0.4em\relax Philadelphia: SIAM, 1 2005.

\bibitem{Sandberg2009}
H.~Sandberg and R.~M. Murray, ``{Model reduction of interconnected linear
  systems},'' \emph{Optimal Control Applications and Methods}, vol.~30, no.~3,
  pp. 225--245, 5 2009.

\bibitem{Wortelboer1994FrequencyConfiguration}
P.~M. Wortelboer and O.~H. Bosgra, ``{Frequency weighted closed-loop order
  reduction in the control design configuration},'' \emph{Proceedings of the
  IEEE Conference on Decision and Control}, vol.~3, no. December, pp.
  2714--2719, 1994.

\bibitem{Vandendorpe2008ModelSystems}
A.~Vandendorpe and P.~Van~Dooren, ``{Model Reduction of Interconnected
  Systems},'' in \emph{Model Order Reduction: Theory, Research Aspects and
  Applications}, 2008, pp. 305--321.

\bibitem{Poort2024Balancing-BasedSystems}
L.~Poort, B.~Besselink, R.~H.~B. Fey, and N.~v.~d. Wouw, ``{Balancing-Based
  Reduction for Interconnected Passive Systems},'' \emph{IEEE Transactions on
  Control Systems Technology}, vol.~32, no.~5, pp. 1817--1826, 2024.

\bibitem{Kessels2022Sensitivity-BasedReduction}
B.~M. Kessels, M.~L.~J. Verhees, A.~M. Steenhoek, R.~H.~B. Fey, and N.~van~de
  Wouw, ``{Sensitivity-Based Substructure Error Propagation for Efficient
  Assembly Model Reduction},'' in \emph{Dynamic Substructures, Volume 4,
  Conference Proceedings of the Society for Experimental Mechanics Series},
  M.~Allen, W.~D'Ambrogio, and D.~Roettgen, Eds.\hskip 1em plus 0.5em minus
  0.4em\relax Springer, 2022, pp. 1--11.

\bibitem{Kim2018ADisplacement}
S.~M. Kim, J.~G. Kim, K.~C. Park, and S.~W. Chae, ``{A component mode selection
  method based on a consistent perturbation expansion of interface
  displacement},'' \emph{Computer Methods in Applied Mechanics and
  Engineering}, vol. 330, pp. 578--597, 2018.

\bibitem{Li2005Structure-PreservingFormulation}
R.-C. Li and Z.~Bai, ``{Structure-Preserving Model Reduction Using a Krylov
  Subspace Projection Formulation},'' \emph{Communications in Mathematical
  Sciences}, vol.~3, no.~2, pp. 179--199, 2005.

\bibitem{Li2005StructuredLMIs}
L.~Li and F.~Paganini, ``{Structured coprime factor model reduction based on
  LMIs},'' \emph{Automatica}, vol.~41, no.~1, pp. 145--151, 2005.

\bibitem{Villena2009BlockSystems}
J.~{Fern{\'a}ndez Villena}, W.~Schilders, and L.~Silveira,
  \emph{\BIBforeignlanguage{English}{Block oriented model order reduction of
  interconnected systems}}, ser. CASA-report.\hskip 1em plus 0.5em minus
  0.4em\relax Technische Universiteit Eindhoven, 2009, vol. 0901.

\bibitem{Leung2019ModelApproach}
J.~Leung, M.~Kinnaert, J.~C. Maun, and F.~Villella, ``{Model reduction in power
  systems using a structure-preserving balanced truncation approach},''
  \emph{Electric Power Systems Research}, vol. 177, no. May, p. 106002, 2019.

\bibitem{Janssen2023ModularApproach}
L.~A.~L. Janssen, B.~Besselink, R.~H.~B. Fey, and N.~van~de Wouw, ``{Modular
  Model Reduction of Interconnected Systems: A Top-Down Approach},''
  \emph{IFAC-PapersOnLine}, vol.~56, no.~2, pp. 4246--4251, 2023.

\bibitem{Ceton1993FrequencyReduction}
C.~Ceton, P.~M.~R. Wortelboer, and O.~H. Bosgra, ``{Frequency weighted closed
  loop balanced reduction},'' \emph{Proceedings of the 2nd European Control
  Conference}, pp. 697--701, 1993.

\bibitem{Wortelboer1994Frequency-weightedTools}
P.~M. Wortelboer, ``{Frequency-weighted balanced reduction of closed-loop
  mechanical servo-systems: theory and tools},'' Ph.D. dissertation, Delft
  University of Technology, 1994.

\bibitem{Zhou1998EssentialsControl}
K.~Zhou and J.~C. Doyle, \emph{{Essentials of Robust Control}}, 1st~ed.\hskip
  1em plus 0.5em minus 0.4em\relax Prentice Hall, 1998.

\bibitem{Cheng2019BalancedSystems}
X.~Cheng, J.~M. Scherpen, and B.~Besselink, ``{Balanced truncation of networked
  linear passive systems},'' \emph{Automatica}, vol. 104, pp. 17--25, 2019.

\bibitem{Moore1981PrincipalReduction}
B.~C. Moore, ``{Principal Component Analysis in Linear Systems:
  Controllability, Observability, and Model Reduction},'' \emph{IEEE
  Transactions on Automatic Control}, vol.~26, no.~1, pp. 17--32, 1981.

\bibitem{Antoulas2005AnSystems}
A.~C. Antoulas, ``{An overview of approximation methods for large-scale
  dynamical systems},'' \emph{Annual Reviews in Control}, vol.~29, no.~2, pp.
  181--190, 2005.

\bibitem{Beck1996ModelSystems}
C.~L. Beck, J.~C. Doyle, and K.~Glover, ``{Model reduction of MultiDimensional
  and Uncertain Systems},'' \emph{IEEE Transactions on Automatic Control},
  vol.~41, no.~10, pp. 1466--1477, 1996.

\bibitem{Gugercin2008H2Systems}
S.~Gugercin, A.~C. Antoulas, and C.~Beattie, ``{H2 Model Reduction for
  Large-Scale Linear Dynamical Systems},'' \emph{SIAM Journal on Matrix
  Analysis and Applications}, vol.~30, no.~2, pp. 609--638, 1 2008.

\bibitem{Craig1981Component-ModeSynthesis}
R.~Craig, ``{Component-Mode Synthesis},'' in \emph{Structural dynamics—an
  introduction to computer methods}.\hskip 1em plus 0.5em minus 0.4em\relax New
  York, NY: Wiley, 1981, ch.~17, pp. 531--575.

\bibitem{Janssen2024ModularPerspective}
L.~A.~L. Janssen, B.~Besselink, R.~H.~B. Fey, and N.~van~de Wouw, ``{Modular
  model reduction of interconnected systems: A robust performance analysis
  perspective},'' \emph{Automatica}, vol. 160, 2024.

\bibitem{Zhou1996RobustControl}
K.~Zhou and J.~C. Doyle, \emph{{Robust and Optimal Control}}.\hskip 1em plus
  0.5em minus 0.4em\relax Prentice Hall, 1996, vol.~4.

\bibitem{Packard1993TheValue}
A.~Packard and J.~C. Doyle, ``{The Complex Structured Singular Value},''
  \emph{Automatica}, vol.~29, no.~1, pp. 71--109, 1993.

\bibitem{Bruinsma1990AMatrix}
N.~A. Bruinsma and M.~Steinbuch, ``{A fast algorithm to compute the
  H$_\infty$-norm of a transfer function matrix},'' \emph{Systems and Control
  Letters}, vol.~14, no.~4, pp. 287--293, 1990.

\bibitem{Glover1984AllBounds}
K.~Glover, ``{All optimal Hankel-norm approximations of linear multivariable
  systems and their L,$\infty$-error bounds†},'' \emph{International Journal
  of Control}, vol.~39, no.~6, pp. 1115--1193, 1984.

\bibitem{Liu1989SingularSystems}
Y.~Liu and B.~D. Anderson, ``{Singular perturbation approximation of balanced
  systems},'' \emph{International Journal of Control}, vol.~50, no.~4, pp.
  1379--1405, 1989.

\bibitem{Gawronski2004AdvancedStructures}
W.~K. Gawronski, \emph{{Advanced Structural Dynamics and Active Control of
  Structures}}.\hskip 1em plus 0.5em minus 0.4em\relax Springer-Verlag, 2004.

\bibitem{Herting1985AMethod}
D.~N. Herting, ``{A general purpose, multi-stage, component modal synthesis
  method},'' \emph{Finite Elements in Analysis and Design}, vol.~1, no.~2, pp.
  153--164, 1985.

\bibitem{mosek}
M.~ApS, \emph{MOSEK Optimization Toolbox for MATLAB 10.1.}, 2023.

\bibitem{Young1993RobustnessUncertainty}
P.~M. Young, ``{Robustness with Parametric and Dynamic Uncertainty},'' Ph.D.
  dissertation, California Institute of Technology, 1993.

\end{thebibliography}

\begin{IEEEbiography}[{\includegraphics[width=1in,height=1.25in,clip,keepaspectratio]{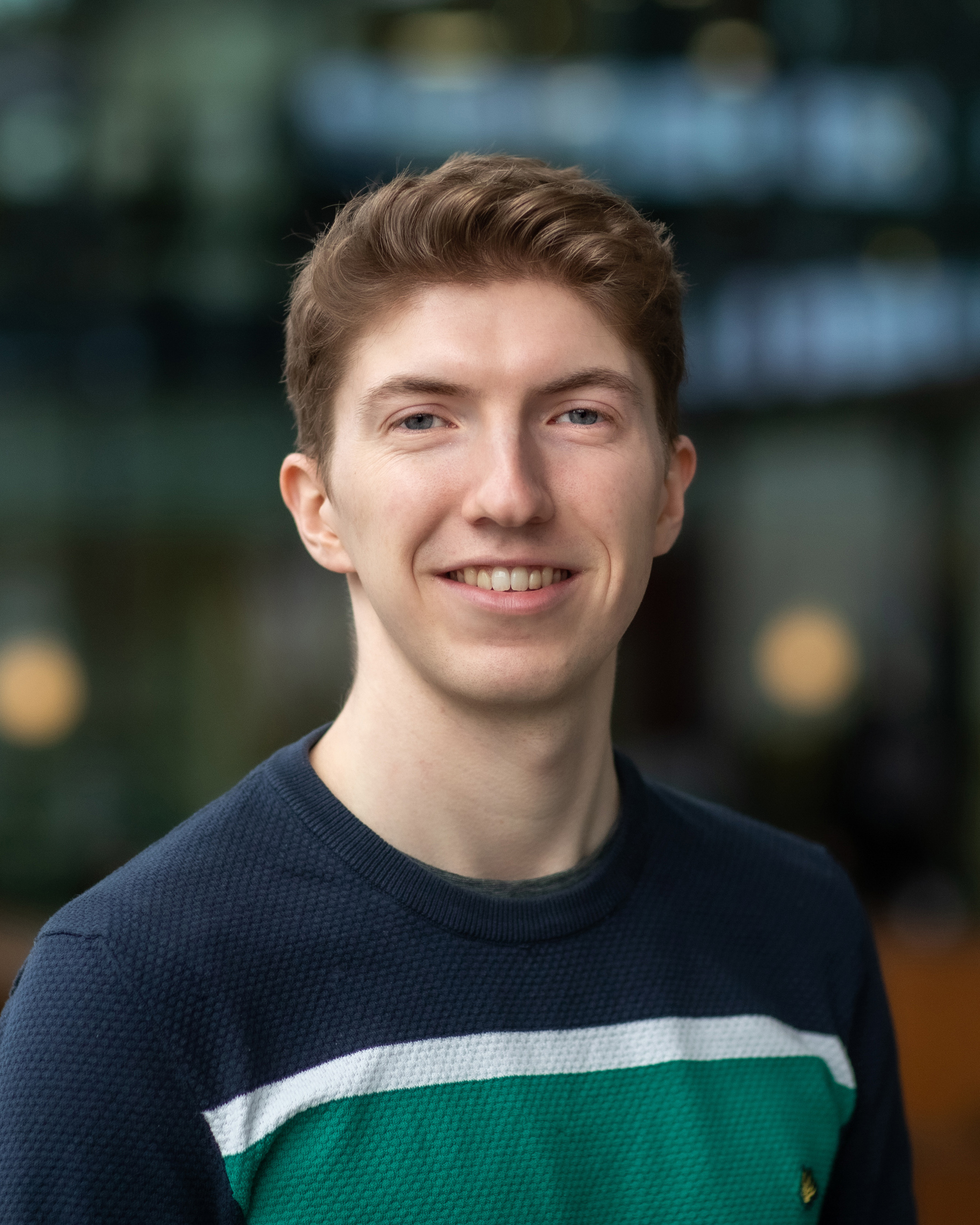}}]{Luuk Poort}
received his M.Sc.-degree (cum laude) in Mechanical Engineering at the Eindhoven University of Technology in 2020. He now works as a doctoral candidate on the derivation of modular model reduction techniques and their application to industrial-scale, structural dynamics models. His research interests include model reduction, structural dynamics and system theory. He is recipient of the best presentation award of the 42nd Benelux meeting on Systems and Control (2023).
\end{IEEEbiography}\vspace{-13mm}
\begin{IEEEbiography}[{\includegraphics[width=1in,height=1.25in,clip,keepaspectratio]{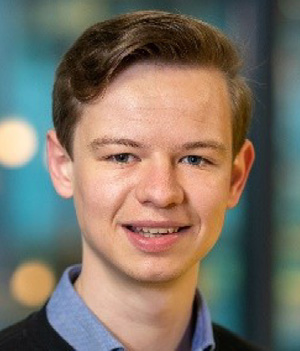}}]{Lars A.L. Janssen}
was born in Nijmegen, the Netherlands,
in 1996. He received his M.Sc.-degree (cum
laude) in Mechanical Engineering from the Eindhoven
University of Technology, Eindhoven, the Netherlands
in 2019. Currently, he is a Doctoral Candidate at the
Dynamics and Control (DC) group of the Mechanical
Engineering Department at Eindhoven University
of Technology (TU/e). His current research interest
are large-scale interconnected systems, modelling of
complex systems and structures, model reduction, and
systems engineering.
\end{IEEEbiography}\vspace{-13mm}
\begin{IEEEbiography}[{\includegraphics[width=1in,height=1.25in,clip,keepaspectratio]{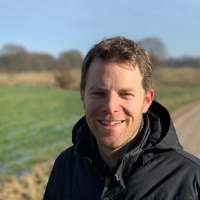}}]{Bart Besselink}
(M’17) received the M.Sc. (cum laude) degree in mechanical engineering in 2008 and the Ph.D. degree in 2012, both from Eindhoven University of Technology, Eindhoven, The Netherlands. Since 2016, he has been an Assistant Professor with the Bernoulli Institute for Mathematics, Computer Science and Artificial Intelligence, University of Groningen, Groningen, The Netherlands. He was a short-term Visiting Researcher with the Tokyo Institute of Technology, Tokyo, Japan, in 2012. Between 2012 and 2016, he was a Postdoctoral Researcher with the ACCESS Linnaeus Centre and Department of Automatic Control, KTH Royal Institute of Technology, Stockholm, Sweden. His main research interests are on mathematical systems theory for largescale interconnected systems, with emphasis on contract-based design and control, compositional analysis, model reduction, and applications in intelligent transportation systems and neuromorphic computing. He is a recipient (with Xiaodong Cheng and Jacquelien Scherpen) of the 2020 Automatica Paper Prize.
\end{IEEEbiography}\vspace{-13mm}
\begin{IEEEbiography}[{\includegraphics[width=1in,height=1.25in,clip,keepaspectratio]{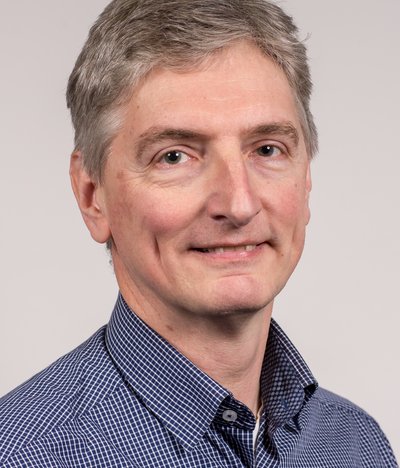}}]{Rob H. B. Fey}
received his M.Sc. degree (cum laude) in Mechanical Engineering and his Ph.D. degree from Eindhoven University of Technology in the Netherlands, in 1987 and 1992, respectively. He was a recipient of the Shell Study Prize for his Ph.D. thesis. From 1992 to 2002, he was a Senior Scientist with the Structural Dynamics Group of the Netherlands Organization for Applied Scientific Research (TNO) in Delft, the Netherlands. Since 2002, he has been with the Dynamics \& Control Group, Department of Mechanical Engineering, Eindhoven University of Technology, where he currently is an Associate Professor of Structural Dynamics. He is (co-)author of many refereed journal papers, chapters in books, and conference papers. His general research interests include the modeling, analysis, and validation of the dynamic behavior of complex structures and (multiphysics) systems. Current focus is on data- and AI-based model updating techniques and model reduction of interconnected systems. His main applications are currently in the fields of High-Tech Systems, Mechatronic Systems and Micro-ElectroMechanical Systems. He is member of the Editorial Board of the Journal of Vibration and Control and the Technical Committee for Vibrations of IFToMM. He was guest editor of the journal Nonlinear Dynamics.
\end{IEEEbiography}\vspace{-13mm}
\begin{IEEEbiography}[{\includegraphics[width=1in,height=1.25in,clip,keepaspectratio]{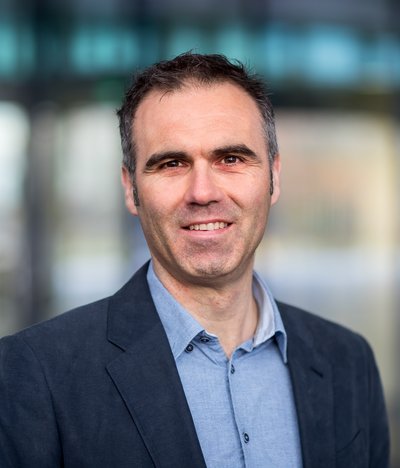}}]{Nathan van de Wouw} obtained his M.Sc.-degree (with honors) and Ph.D.-degree in Mechanical Engineering from the Eindhoven University of Technology, Eindhoven, the Netherlands, in 1994 and 1999, respectively. He currently holds a full professor position at the Mechanical Engineering Department of the Eindhoven University of Technology, the Netherlands. His current research interests are the modeling, model reduction, analysis and control of nonlinear/hybrid and delay systems, with applications to vehicular platooning, high-tech systems, resource exploration, smart energy systems and networked control systems.\end{IEEEbiography}
\end{document}